\documentclass[epj]{svjour}
\usepackage{graphics,figsize}

\begin{document}
\title{Non-extensive Statistics and a Systematic Study of Meson-Spectra at LHC Energy $\sqrt{s_{NN}}=2.76$ TeV}

\author{Bhaskar De\thanks{e-mail: bhaskar.de@gmail.com}}

\institute{Department of Physics\\
A. P. C. Roy Government College\\
Himachal Bihar, Matigara, Siliguri-734010\\
West Bengal, India.\\
}

\date{}
%\maketitle

\abstract{
The transverse momentum spectra of secondary pions and kaons, produced in $P+P$ and various central $Pb+Pb$ collisions at $\sqrt{s_{NN}}=2.76$ TeV at LHC, have been analyzed systematically with an approach based on Tsallis non-extensive statistics. The analytical results have been utilized to determine some of the very important thermodynamical parameters bearing characteristic signatures of the partonic medium produced in such collisions.
 % end of PACS codes
}

\authorrunning{B. De}
\titlerunning{Non-extensive Statistics and a Systematic Study at $\sqrt{s_{NN}}=2.76$ TeV}
\maketitle

\newpage
%\doublespacing
\section{Introduction}
The multiparticle production rate in ultrarelativistic heavy ion collision experiments and their momentum distributions is very sensitive to and carry useful information on possible occurrence of a phase transition from a deconfined partonic state, produced due to deposition of enormous amount of energy in the vicinity of the centre of mass of two colliding beams, to confined hadronic sates. The analyses of particle-spectra with a suitable model/approach, thus, pave the way to get insights of the thermodynamical evolution of such a deconfined medium and it's cooling off through the process of hadronization. Analysis of such a system is statistical in nature; and Tsallis generalized non-extensive statistics, instead of usual Boltzmann-Gibbs statistics, has been proven, over the years, to be a good choice to deal with such a system\cite{Tsallis1,Tsallis2,Tsallis3,Tsallis4,Tsallis5,Prato1,Beck1,Beck2,Beck3,Wilk1,Wilk1.1,Wilk1.2,Wilk1.3,Wilk1.4,Wilk2,Wilk3,Wilk4,Wilk5,Osada1,Biro1,Biro2,Biro3,Biro3.1,Biro4,Urmossy0.1,Biyajima1,Biyajima2,Alberico1,Lavagno1,Kaniadakis,Kodama,De1,De2,Wibig1,Wibig2,Jiulin1,Jiulin2,Urmossy1,Urmossy2}. The mechanism behind the emergence of Tsallis-like spectra from such a hot and dense partonic matter, produced immediate after the nuclear interactions at ultrarelativistic energies, is yet to be understood theoretically. However, the possible reasons could be as follows: (i) The produced fireball, consisting the partons, may not be fully deconfined or weakly-coupled; but strong correlations or $N$-body interactions among its costituents may exist\cite{Shuryak1,Hirano1}. The presence of such long-range correlations may give rise to the non-Markovian nature of the hadronizing system and hence, to Tsalis-spectra\cite{Biro1,Biro2,Biro3,Biro3.1,Biro4,Urmossy0.1}. (ii) Besides, the particle-spectra from such nuclear interactions are, generally, obtained by averaging, statistically, over million of events; and the appearance of power-law-like tail in hadron-spectra, which is reproduced, quite successfully, by Tsallis generalised statistics, may also be generated due to event-by-event fluctuations\cite{Beck1,Beck2,Beck3,Wilk1,Wilk1.1,Wilk1.2,Wilk1.3,Wilk1.4,Wilk2,Wilk3,Wilk4,Wilk5,Urmossy1} of various observables\cite{Jena1,Koch1,Koch2,Bialas1} characterising such a hot and dense medium.
\par
Eversince, the data on different hadronic-spectra have started pouring in from LHC experiments, different derivatives of Tsallis non-extensive statistics is in extensive use to interpret different aspects of hadronizing medium by various theoretical groups\cite{Urmossy1,Urmossy2,Deppman1,Deppman2,Deppman3,Deppman4,Deppman5,Cleymans1,Cleymans2,Cleymans3,Cleymans4,Cleymans5,Rybczynski1,Rybczynski2}. Most of these groups, so far, confined their studies, mainly, to the $P+P$ collisions at all the colliding energies at LHC.  In Ref.\cite{Deppman1,Deppman2,Deppman3,Deppman4,Deppman5,Cleymans1,Cleymans2,Cleymans3,Cleymans4,Cleymans5}, analyses of transverse-momentum($p_T$)-spectra of charged hadrons alongwith some identified ones produced in $P+P$ collisions  at LHC and in different central nuclear interactions at RHIC energies  were done by developing a `self-consistent' theory on the basis of non-extensive statistics; whereas in Ref.\cite{Urmossy1,Urmossy2}, particle-spectra as a function of $p_T$, energy fraction and longitudinal momentum fraction for a fixed event-multiplicity were under scanner with the help of another version  called `super-statistics'. However, in Ref.\cite{Rybczynski1,Rybczynski2}, charged hadron spectra from $Pb+Pb$ interaction at 2.76 TeV at LHC, alongwith those produced in $P+P$, $P+\bar{P}$ and different $A+A$ interactions over a wide range of colliding energies, were taken into account to find out a possible scaling behaviour, called `$q$-scaling', of the non-extensivity parameter $q$ extracted from analyses of all the experimental data at different  energies.     
\par
The present author alongwith his collaborators had made efforts\cite{De1,De2} to study the impact of Tsallis non-extensive statistics on some of the identified hadronic-spectra available from RHIC experiments following the phenomenological prescriptions made in Ref.\cite{Beck1,Wilk3}.
In the present work, a similar task of systematic analyses of the meson-spectra, available from LHC experiments, has been taken up. Such a systematic study demands to have data for a particular variety, to be analysed, for all the centrality bins of colliding $Pb$ nuclei in addition to the same for $P+P$ collisions at all the colliding energies. But, till this date, a moderate range of data is available only for $\sqrt{s_{NN}}=2.76$ TeV. So, for the present study, we concentrate mainly on a single colliding energy at LHC with two abundant light mesonic varieties --- $\pi$- and $K$-mesons. Beside this, the parameters obtained from the analyses have also been used to determine some of the thermodynamical parameters, such as pressure, energy density, trace anomaly, square of velocity of sound etc. which provide useful information on the equation of state of the fireball produced in ultrarelativistic nuclear collisions; and the results have further been compared with a Lattice-QCD-based calculation.
\par
The work is organized as follows: Section 2 presents a brief
sketch of the nonextensive statistics and the main working formulae
to be used in the present study, here.  The obtained results are reported in next
section(Section 3) with some specific observations made. And the
last section is preserved for the concluding remarks.

\section{Nonextensive Statistics and Transverse Momentum Spectra}

The generalized statistics of Tsallis is not only applicable to an
equilibrium system, but also to nonequilibrium systems with
stationary states\cite{Beck2}. As the name `nonextensive' implies,
these entropies are not additive for independent systems.
\par
The nonextensive Boltzmann factor is defined as\cite{Beck2}
\begin{equation}
x_{ij} ~ = ~ (1 ~ + ~ (q-1)\beta \epsilon_{ij})^{-q/(q-1)}
\end{equation}

 where $\epsilon_{ij} ~ = ~ \sqrt{\textbf{p}_i^2 ~ + ~ m_j^2}$ is the energy associated with $j-$th particle of rest mass $m_j$ in momentum state $i$,  $\beta=1/T$ is the inverse temperature variable, $q$ is a measure of degree of fluctuation present in the system and is called nonextensivity parameter; with $q\rightarrow1$, the above equation approaches the ordinary Boltzmann factor $e^{-\beta\epsilon_{ij}}$.
\par
 If $\nu_{ij}$ denotes
the number of particles of type $j$ in momentum state $i$, the
generalized grand canonical partition function is given by,

\begin{equation}
Z ~ = ~ \sum_{(\nu)} \prod_{ij} x_{ij}^{\nu_{ij}}
\end{equation}

The average occupation number of a particle of species $j$ in the
momentum state $i$ can be written as\cite{Beck2}

\begin{equation}
{\bar{\nu}_{ij}}= x_{ij}\frac{\partial}{\partial x_{ij}} \log{Z} ~
= ~ \frac{1}{(1+(q-1)\beta\epsilon_{ij})^{q/(q-1)}\pm 1}
\end{equation}

where $-$ sign is for bosons and the $+$ sign is for fermions.\\
The probability of observation of a particle of mass $m_0$ in a certain
momentum state can be obtained by multiplying the average occupation number with
the available volume in momentum space\cite{Beck2}. The infinitesimal volume in momentum space
is given by

\begin{equation}
d^3p ~ = ~ E ~ dy ~ p_T ~ dp_T ~ d\phi
\end{equation}

where $E$ is the energy, $p_T$ is the transverse momentum and $y$ is the rapidity and $\phi$ is the azimuthal angle.
Hence, the probability density $w(p_T,y,\phi)$  is given by:

\begin{equation}
w(p_T,y,\phi) ~ \propto ~ 
\frac{1}{(1+(q-1)\beta E)^{q/(q-1)} \pm 1} ~  E ~ dy ~ p_T ~ dp_T ~ d\phi
\end{equation}

Assuming azimuthal symmetry, one would obtain the invariant distribution as
\begin{equation}
\frac{1}{2\pi} ~ \frac{d^2N}{p_Tdp_Tdy}  ~ = ~ C ~ 
\frac{E}{[1+(q-1)\beta E]^{q/(q-1)} \pm 1}
\end{equation}
where $C$ is a proportionality constant.\\

Using the relationships $\beta = \frac{1}{T_{eff}}$ and $E=m_T ~ coshy$, where $T_{eff}$ is the effective temperature of the interaction region and $m_T=\sqrt{m_0^2+p_T^2}$ is the transverse mass, the invariant yield at mid-rapidity(for $y\simeq 0$ ) will take the form

\begin{equation}
\frac{1}{2\pi} ~ \frac{d^2N}{p_Tdp_Tdy}  ~ = ~ C ~ 
\frac{m_T}{[1+(q-1) ~ \frac{m_T}{T_{eff}}]^{q/(q-1)} \pm 1}
\end{equation}
\par
The average multiplicity of the detected secondary per unit rapidity in the given rapidity region can be obtained by the relationship
\begin{equation}
\begin{array}{lcl}
\frac{dN}{dy}  & = &  \int_0^\infty \frac{d^2N}{dp_T ~ dy} ~ dp_T  \\
& & \\
 & = & ~ C_1 ~\int_0^\infty \frac{m_T}{[1+(q-1) ~ \frac{m_T}{T_{eff}}]^{q/(q-1)} \pm 1} p_T dp_T
\end{array}
\end{equation}
 
where $C_1=2\pi C$.\\
Hence, the constant $C_1$ can be expressed in terms of $\frac{dN}{dy}$ by the relationship

\begin{equation}
C_1 = \frac{dN}{dy} ~ \frac{1}{\int_0^\infty \frac{m_T}{[1+(q-1) ~ \frac{m_T}{T_{eff}}]^{q/(q-1)} \pm 1} p_T dp_T}
\end{equation}

\par
Combination of eqn(7) and eqn(9) will provide us the main working formula for invariant yield for a detected secondary and it is given by
\begin{equation}
\begin{array}{lcl}
 \frac{d^2N}{p_T ~ dp_T ~ dy} ~ & = & ~ \frac{dN}{dy} ~ \frac{1}{\int_0^\infty \frac{m_T}{[1+(q-1) ~ \frac{m_T}{T_{eff}}]^{q/(q-1)} \pm 1} p_T dp_T} \\
 
& & \times  \frac{m_T}{[1+(q-1) ~ \frac{m_T}{T_{eff}}]^{q/(q-1)} \pm 1}
 \end{array} 
\end{equation}

\par
Further, it was observed earlier that the parameters $T_{eff}$ and $q$ are strongly correlated, even if they set free\cite{Wilk3,De1}. So, these two parameters alongwith average multiplicity can phenomenologically be correlated  
by the following relationships\cite{Wilk3}:

\begin{equation}
T_{eff}=T_0(1-c(q-1))
\end{equation}

\begin{equation}
\frac{<N> \sim n_0N_{part}}{<N>}=c(q-1)
\end{equation}

with $<N>=\frac{dN}{dy}$ and $c= - \frac{\phi}{D c_p\rho T_0}$ where $D$, $c_p$, $\rho$,
$T_0$ are respectively the strength of the temperature fluctuations,
the specific heat under constant pressure, density, the critical temperature(also called the Hagedorn temperature\cite{Hagedorn1,Hagedorn2})
of the hadronizing system when it is in thermal equilibrium($q=1$)
and $N_{part}$ is the number of participant nucleons. Eqn.(11)
describes the fluctuation in temperature where it is assumed that
the effective temperature $T_{eff}$ is the outcome of two
simultaneous processes: (i) the fluctuation of the temperature
around $T_0$ due to a stochastic process in any selected region of
the system and (ii) some energy transfer between the selected region
and the rest of the system, denoted by $\phi$\cite{Wilk3}. It is
absolutely uncertain whether such energy-transfers could/should be
invariably linked up with flow-velocity(normally denoted in the
hydrodynamical model-texts as `$u$'). So, for the sake of
calculational simplicity and correctness we assume the factor
$\phi$, for the present, to be independent of any flow-velocity. The
fluctuation in multiplicity is described by eqn.(12). The assumption
behind this relationship is that if N-particles are distributed in
energy according to Tsallis non-extensive distribution, then their
multiplicity will obey Negative-Binomial distribution\cite{Wilk3}. In the original work\cite{Wilk3}, a `$-$' sign was used in place of `$\sim$' on the left-hand side of equation(12). But, the present modified form of equation(12) will take into account only the magnitude of difference between the product $n_0N_{part}$ and $<N>$, even if the product term exceeds the average term, and hence, will keep $c$ positive and, in turn, $\phi$ negative ensuring that the energy is transferred from the interaction region to the spectators of the non-interacting nucleons\cite{Wilk3}.
\par
Equation(10) alongwith the constraints imposed by equations (11-12) provides the working formula for the present analysis.
\par
Once, the parameters $T_{eff}$ and $q$ are in hand, the logarithm of partition function can numerically be calculated by the relationship\cite{Beck1}:

\begin{equation}
\begin{array}{lcl}
\frac{1}{V} logZ ~ & = & ~ \frac{1}{(2\pi)^3}[\int d^3p \int_{m_\pi}^M dm ~ \rho(m) log(\frac{1}{1-x_{ij}}) \\
& & + \int d^3p \int_{m_P}^M dm ~ \rho(m) log(1+x_{ij})]
\end{array}
\end{equation}  

where the first term inside the parenthesis belongs to mesons and the second one for baryons. $V$ is the volume of interaction region and $\rho(m)$ is the hadronic mass spectrum. The parametrization for $\rho(m)$ used here is given by\cite{Deppman4}, 

\begin{equation}
\rho(m) ~ = ~ \gamma m^{-5/2}[1+(q_0-1)\frac{m}{\tau_0}]^{\frac{1}{q_0 -1}}
\end{equation}    

with the parameter-values given by $\gamma = 5 \times 10^{-3}$ GeV$^{3/2}$, $\tau_0=0.607$ GeV and $q_0 = 1.138$. This parametrization is suitable upto $m=2.5$ GeV. So, the upper limit of integration with respect to $m$ has been kept $M=2.5$ GeV\cite{Deppman4}; while the lower limit for mesonic-part  is the pion mass($m_\pi=0.140$ GeV) and that for baryons is the mass of proton($m_P = 0.938$ GeV). A point is to be noted here that the above parametrization is used here only to avail a continuous description of the hadron mass-spectrum over the specified  region of hadronic-mass.

Now, it is possible to determine, numerically, the values of the following transport coefficients\cite{Deppman4} which are essential entities to determine the equation of state of the matter formed after nuclear interactions at very high energies:   

\begin{equation}
\Pi = \frac{T_{eff}}{V} logZ
\end{equation}    

\begin{equation}
s = \frac{\partial \Pi}{\partial T_{eff}}
\end{equation}    

\begin{equation}
\epsilon = \frac{T_{eff}^2}{V} \frac{\partial logZ}{\partial T_{eff}}
\end{equation}    

\begin{equation}
\alpha = \frac{\epsilon - 3\Pi}{T_{eff}^2}
\end{equation}    

\begin{equation}
c_V = \frac{\partial \epsilon}{\partial T_{eff}}
\end{equation}    

\begin{equation}
c_s^2 = \frac{s}{c_V}
\end{equation}    

where $\Pi$, $s$, $\epsilon$, $\alpha$, $c_V$, $c_s^2$ are the pressure of the interaction volume, entropy density, energy density, trace-anomaly, specific heat at constant volume and square of the velocity of sound respectively.

\section{Results and Discussions}

The working formula(eqn.(10)) was applied, in it's present form, to obtain the fits to the
data on $\pi^{0,\pm}$ and $K^{\pm}$ production in $P+P$ collisions[Fig.1] at $\sqrt{s_{NN}}=2.76$ TeV; and
the corresponding parameter-values are given in Table-1, where $n_o$ denotes the average
multiplicity per unit rapidity of the produced meson-variety in $P+P$ interaction.
\par
The fits for different central $Pb+Pb$ collisions obtained on the basis of
equation(10) alongwith the constraints given in eqn(11)-(12) are
depicted in Fig.2 \& in Fig.3. The values of various parameters 
obtained from the fits are provided in Table-2-Table-3. It is observed from the values of $\chi^2/ndf$, enlisted in Table-1-Table-3, that the performance of the present approach is quite satisfactory in reproducing the experimental data, except the cases for pion production at $80-90\%$ central collisions.  
\par
The values of the Hagedorn's temperature($T_0$) obtained from the fits for various centralities have been depicted graphically in Fig.4, which exhibits almost constant behaviour for a particular secondary-type emitted in the nuclear interactions at LHC energy 2.76 TeV. The average value of $T_0$ is found to be $0.144\pm 0.002$ GeV from the analysis of pion-spectra which is in good agreement with the findings from analysis of pion production in $e^+e^-$ collisions($T_0=0.131$ GeV)\cite{Wilk4}. On the otherhand, the same, obtained by analysing kaon-spectra, has the average value $0.222\pm 0.001$ GeV which lies in the close vicinity of the predictions($0.170-0.200$ GeV) made by Lattice-QCD calculations\cite{Ejiri1,Miura1} and the result($0.192 \pm 0.015$ GeV) obtained from a recent analysis of hadron-spectra from $P+P$ collisions over a wide range of energies\cite{Deppman3}.  
\par
In our earlier analysis\cite{De2} of $\eta$-spectra at RHIC energies, the parameter, $c$, also exhibited a constant behaviour with an average value 1.880. However, in the present study, leaving two or three cases, in general, an increasing trend is observed for $c$ while going from central to peripheral interactions.  
\par
The effective temperature, $T_{eff}$,
and the non-extensive parameter, $q$, calculated from the fitted parameters( excluding the errors), are given in tabular form in Table-4 and in graphical format in Fig.5 as a function of
participant nucleons. Excluding the results from P+P interactions, $T_{eff}$ decreases while $q$ increases for both the varieties, in general, when going from $N_{part}=382.8$\cite{Abelev2,Aamodt1,Abelev3} for $0-5\%$ central to $N_{part}=7.5$ for $80-90\%$ central $Pb+Pb$ collision at 2.76 TeV. A deviation, though very weak, from this trend is observed for K-meson production in the central collision regions with $N_{part}\geq 100$.  

\par 
Fig.6(a) depict graphically the behaviour of
average  multiplicity($<N>=\frac{dN}{dy}$) of the detected secondary with respect to $N_{part}$ while Fig.6(b-c) represent the same, but this time normalized by per pair of participant nucleons( $N_{part}/2$) and by pair of participant quarks( $N_{q-part}/2$) respectively. The values of $N_{part}$ for different centralities have been obtained from Ref.\cite{Abelev2,Aamodt1,Abelev3} and those of $N_{q-part}$(Table-4) have been calculated using PHOBOS Glauber Monte Carlo Simulation\cite{Alver1} with incorporation of the prescription made in Ref.\cite{Voloshin1} and the method employed in Ref.\cite{Netrakanti1,De3} into the code. 
A close inspection of Fig.6(b) and Fig.6(c) will reveal that the dependence of the normalized 
yield of both the varieties on $N_{q-part}$  is not so prominent compared to that on  $N_{part}$ for most of the centrality-bins which indicates a nearly linear dependence of $<N>$ on $N_{q-part}$. The similar observation was made for meson-production at RHIC interactions\cite{De3}. 
\par
In Fig.7-Fig.8, the nature of various transport coefficients, calculated numerically on the basis of eqn.(15)-eqn.(20) and obtained for different centrality classes of $Pb+Pb$ interactions, have been represented graphically. One of the useful parameters, square of the velocity of sound($c_s^2$), was calculated from rapidity spectra in four most central $Pb+Pb$ collisions at $\sqrt{s_{NN}}=2.76$ TeV in Ref.\cite{Gao1}. These results have also been incorporated in Fig.8(b) to provide a comparison with the results obtained from the present analysis.
\par 
Fig.9 is the graphical comparison of our results on pressure, trace anomaly and square of velocity of sound as functions of effective temperature( $T_{eff}$) with those obtained by a calculation on the basis of 
Lattice-QCD-Thermodynamics\cite{Borsanyi1} with zero chemical potential. It is observed that the values of all the three entities, calculated using the parameters obtained from the fits of both the mesonic-spectra, are in moderate agreement with the Lattice-QCD-based calculations at high temperature region. Moreover, the values of square of the velocity of sound  remains nearly constant over entire range of effective temperature. Accumulation of multiple data in the close vicinity of a particular temperature is due to variation in the value of $q$.

\section{Conclusions}
As $T_{eff}$ and $q$ are mutually correlated and exhibit strong contrast behaviour, one expects that an increase in temperature of the interaction volume will be exhibited by very low value of $q$, i.e., the interaction region will have lesser degree of fluctuations, and will be in the vicinity of thermal equilibrium. The values of effective temperature of the interaction volume, obtained from the analysis of $\pi$-spectra for different centralities, in the present study, are found to be somewhat similar compared to those observed for RHIC interactions\cite{De1,De2}. However, the same parameter is found to be a bit high for $K$-spectra. But, surprisingly, there is no significant decrement in the associated value of $q$, obtained from kaon-spectra at a particular centrality, though the corresponding effective temperature is quite high compared to that obtained for pion-spectra. Besides, the values of $q$ remain almost same with respect to those observed at RHIC energies\cite{De1,De2} at different centralities. 
This observation indicates that either the fireball, produced in nuclear collisions at this particular LHC energy, possesses partonic constituents which are correlated mutually through long-range interactions, or the presence of event-by-event fluctuations in the particle-production rate. This fact is once again validated from Fig.9 where the obtained values of both the parameters --- the trace anomaly and the square of the velocity of sound --- are far away from their respective ideal gas limits $\frac{\epsilon - 3\Pi }{T^4} \rightarrow 0$ and $c_s^2\rightarrow \frac{1}{3}$. 
\par
Hence, from various signatures available from present analyses, like (i) the value of Hagedorn temperature($T_{0} \sim m_{\pi}$), (ii) the linear-dependence of $<N>$ on $N_{q-part}$, (iii) presence of long-range correlations and/or multiplicity fluctuations,  etc., there are ample reasons to assume that the process of hydrodynamic evolution of the fireball and it's subsequent hadronization is quite similar to that of the processes observed at RHIC energies. This inference is in accord with the observations made in Ref.\cite{Shuryak1}.  
\par
However, one point is to be noted here that we have not incorporated any type of collective transverse flow in our present approach to extract information from the transverse momentum spectra, where, mainly, light mesons have been dealt with. The influence of such collective flow becomes more significant for production of heavier hadrons as it contributes more to the average transverse momentum, associated with a particular secondary, with increment in the mass of the produced hadron keeping the average thermal momentum same for all the varieties\cite{Lee1,Blaizot1}. So, it is quite clear that, in our future endeavour, the present approach may need to be modified, by taking into account the effect of transverse flow, to deal with heavier hadronic-spectra to obtain valuable insights on the thermodynamical evolution of the hot and dense partonic matter produced in nuclear interactions at LHC energies .    
\par
\begin{acknowledgement}
The author would like to express his thankful gratitude to the anonymous referee for his/her valuable suggestions for the improvement of an earlier draft of the manuscript. It is also a great pleasure on his part to thank Prof. S. Bhattacharyya  for some useful discussion while the work was in progress.
\end{acknowledgement}
%\newpage

\newpage
\begin{table*}[h]
\caption{Values of fitted parameters with respect to the experimental
data on meson-spectra produced in $P+P$ collision at $\sqrt{s_{NN}}=2.76$ TeV}
\centering
\begin{tabular}{llllll}
\hline  Meson type & $N_{part}$ & $n_0$ & $q$ & $T_{eff}$(GeV) & $\chi^2/ndf$\\
\hline\noalign{\smallskip}
$\pi^0$ &  & $1.914\pm0.003$  & $1.150\pm0.003$ & $0.078\pm0.004$ & $8.335/16$ \\
%\hline
$\pi^+$ &  & $1.913\pm0.006$  & $1.150\pm0.004$ & $0.078\pm0.003$ & $31.915/36$ \\
%\hline
$\pi^-$ & 2 & $1.913\pm0.005$  & $1.151\pm0.003$ & $0.078\pm0.003$ & $33.714/36$ \\
%\hline
$K^+$ &  & $0.244\pm0.002$  & $1.146\pm0.004$ & $0.089\pm0.005$ & $16.643/36$ \\
%\hline
$K^-$ &  & $0.244\pm0.004$  & $1.146\pm0.003$ & $0.089\pm0.004$ & $14.336/36$ \\
\noalign{\smallskip}\hline\noalign{\smallskip}
\end{tabular}
\end{table*}

\begin{table*}[h]
\caption{Values of fitted parameters with respect to the experimental
data on pion-spectra at different centralities of $Pb+Pb$ collisions at LHC energy $\sqrt{s_{NN}}=2.76$ TeV}
\centering
\begin{tabular}{lllllll}
\hline\noalign{\smallskip} Meson type & Centrality & $N_{part}$ & $<N>$ & $c$ & $T_0$(GeV) & $\chi^2/ndf$\\
\noalign{\smallskip}\hline\noalign{\smallskip}
 & 0-20 & 308 & $499\pm 13$  & $1.581\pm0.002$ & $0.145\pm0.003$ & $2.370/5$ \\
$\pi^0$  & 20-40 & 157 & $248\pm 8$  & $1.773\pm0.002$ & $0.144\pm0.002$ & $3.291/5$ \\
 & 40-60 & 69 & $105\pm 2$  & $2.045\pm0.005$ & $0.144\pm0.003$ & $1.841/5$ \\
 & 60-80 & 23 & $32\pm 1$  & $2.689\pm0.005$ & $0.141\pm0.002$ & $4.277/5$ \\
\noalign{\smallskip}\hline\noalign{\smallskip}
 & 0-5 & 382.8 & $755\pm 72$  & $0.331\pm 0.006$ & $0.145\pm0.007$ & $53.156/38$ \\
& 5-10 & 329.7 & $651\pm 23$  & $0.354\pm0.002$ & $0.143 \pm0.002$ & $72.973/38$ \\
 & 10-20 & 260.5 & $430\pm 12$  & $1.385\pm0.004$ & $0.143\pm0.002$ & $25.218/38$ \\
$\pi^+$ & 20-30 & 186.4 & $300 \pm 6$  & $1.587\pm 0.003$ & $0.145\pm0.002$ & $28.783/38$ \\
 & 30-40 & 128.9  & $210\pm 15$  & $1.60\pm0.03$ & $0.145\pm0.002$ & $51.502/38$ \\
 & 40-50 & 85.0 & $120\pm 6$  & $2.437\pm0.003$ & $0.145\pm0.007$ & $19.514/38$ \\
 & 50-60 & 52.8 & $70\pm 2$  & $2.781\pm0.005$ & $0.143\pm0.003$ & $19.268/38$ \\
 & 60-70 & 30.0 & $38\pm 2$  & $3.036\pm0.003$ & $0.143\pm 0.005$ & $24.912/38$ \\
 & 70-80 & 15.8 & $19.5\pm 0.8$  & $3.37\pm0.02$ & $0.15\pm 0.01$ & $51.925/38$ \\
  & 80-90 & 7.5 & $8.4\pm 0.7$  & $4.02\pm0.02$ & $0.18\pm 0.02$ & $107.797/38$ \\
\noalign{\smallskip}\hline\noalign{\smallskip}
 & 0-5 & 382.8 & $755\pm 63$  & $0.331\pm 0.005$ & $0.145\pm0.008$ & $48.199/38$ \\
& 5-10 & 329.7 & $651\pm 17$  & $0.354\pm0.002$ & $0.143 \pm0.002$ & $67.538/38$ \\
 & 10-20 & 260.5 & $429\pm 11$  & $1.385\pm0.005$ & $0.143\pm0.003$ & $21.182/38$ \\
$\pi^-$ & 20-30 & 186.4 & $300 \pm 11$  & $1.587\pm 0.002$ & $0.145\pm0.002$ & $24.163/38$ \\
 & 30-40 & 128.9  & $210\pm 12$  & $1.60\pm0.02$ & $0.145\pm0.002$ & $47.613/38$ \\
 & 40-50 & 85.0 & $120\pm 8$  & $2.451\pm0.003$ & $0.145\pm0.007$ & $13.621/38$ \\
 & 50-60 & 52.8 & $70\pm 2$  & $2.783\pm0.003$ & $0.143\pm0.004$ & $19.420/38$ \\
 & 60-70 & 30.0 & $38\pm 2$  & $3.036\pm0.003$ & $0.143\pm 0.006$ & $27.799/38$ \\
 & 70-80 & 15.8 & $19.5\pm 0.7$  & $3.32\pm0.01$ & $0.15\pm 0.01$ & $52.373/38$ \\
  & 80-90 & 7.5 & $8.4\pm 0.7$  & $3.96\pm0.05$ & $0.18\pm 0.02$ & $112.440/38$ \\
\noalign{\smallskip}\hline\noalign{\smallskip}
\end{tabular}
\end{table*}
\begin{table*}[h]
\caption{Values of fitted parameters with respect to the experimental
data on kaon-spectra at different centralities of $Pb+Pb$ collisions at LHC energy $\sqrt{s_{NN}}=2.76$ TeV}
\centering
\begin{tabular}{lllllll}
\hline\noalign{\smallskip} Meson type & Centrality & $N_{part}$ & $<N>$ & $c$ & $T_0$(GeV) & $\chi^2/ndf$\\
\noalign{\smallskip}\hline\noalign{\smallskip}
 & 0-5 & 382.8 & $107\pm 9$  & $1.487\pm 0.008$ & $0.222\pm 0.008$ & $36.611/33$ \\
& 5-10 & 329.7 & $91\pm 4$  & $1.335\pm0.006$ & $0.218 \pm0.003$ & $22.858/33$ \\
 & 10-20 & 260.5 & $69\pm 3$  & $1.032\pm0.008$ & $0.220\pm0.005$ & $11.637/33$ \\
$K^+$ & 20-30 & 186.4 & $48 \pm 3$  & $0.734\pm 0.003$ & $0.220\pm0.004$ & $9.363/33$ \\
 & 30-40 & 128.9  & $33\pm 2$  & $0.733\pm0.005$ & $0.217\pm0.006$ & $31.073/33$ \\
 & 40-50 & 85.0 & $18\pm 2$  & $1.827\pm0.003$ & $0.220\pm0.007$ & $1.148/33$ \\
 & 50-60 & 52.8 & $10.5\pm 0.4$  & $2.524\pm0.004$ & $0.220\pm0.008$ & $5.213/33$ \\
 & 60-70 & 30.0 & $5.4\pm 0.3$  & $3.183\pm0.006$ & $0.220\pm 0.007$ & $8.672/33$ \\
 & 70-80 & 15.8 & $2.5\pm 0.2$  & $3.809\pm0.003$ & $0.225\pm 0.002$ & $16.688/33$ \\
  & 80-90 & 7.5 & $1.01\pm 0.05$  & $4.15\pm0.06$ & $0.22\pm 0.01$ & $54.042/33$ \\
\noalign{\smallskip}\hline\noalign{\smallskip}
  & 0-5 & 382.8 & $107\pm 8$  & $1.427\pm 0.005$ & $0.222\pm 0.002$ & $25.271/33$ \\
& 5-10 & 329.7 & $91\pm 3$  & $1.387\pm0.005$ & $0.218 \pm0.003$ & $23.098/33$ \\
 & 10-20 & 260.5 & $69\pm 5$  & $1.032\pm0.006$ & $0.220\pm0.005$ & $8.910/33$ \\
$K^-$ & 20-30 & 186.4 & $48 \pm 3$  & $0.734\pm 0.003$ & $0.220\pm0.005$ & $8.016/33$ \\
 & 30-40 & 128.9  & $33\pm 2$  & $0.733\pm0.004$ & $0.217\pm0.004$ & $30.770/33$ \\
 & 40-50 & 85.0 & $18\pm 2$  & $1.827\pm0.003$ & $0.220\pm0.006$ & $1.368/33$ \\
 & 50-60 & 52.8 & $10.5\pm 0.5$  & $2.524\pm0.006$ & $0.220\pm0.008$ & $5.213/33$ \\
 & 60-70 & 30.0 & $5.3\pm 0.2$  & $3.252\pm0.005$ & $0.220\pm 0.004$ & $6.843/33$ \\
 & 70-80 & 15.8 & $2.5\pm 0.2$  & $3.809\pm0.004$ & $0.225\pm 0.002$ & $19.972/33$ \\
  & 80-90 & 7.5 & $1.01\pm 0.04$  & $4.15\pm0.06$ & $0.22\pm 0.01$ & $42.726/33$ \\
\noalign{\smallskip}\hline\noalign{\smallskip}
\end{tabular}
\end{table*}
\begin{table*}[h]
\caption{Values of $q$ and $T_{eff}$ obtained from different meson-spectra produced at various centralities.}
\centering
\begin{tabular}{llllllllllll}
\hline \noalign{\smallskip}
$N_{part}$ & $N_{q-part}$ & $T_{eff}^{\pi^0}$ & $q^{\pi^0}$ & $T_{eff}^{\pi^+}$  & $q^{\pi^+}$ & $T_{eff}^{\pi^-}$ & $q^{\pi^-}$ & $T_{eff}^{K^+}$ & $q^{K^+}$ & $T_{eff}^{K^-}$ & $q^{K^-}$\\
\noalign{\smallskip}\hline\noalign{\smallskip}
2 & 3.05 & 0.078 & 1.150 & 0.078 & 1.150 & 0.078 & 1.151 & 0.089 & 1.146 & 0.089 & 1.146 \\ 
7.5 & 8 &  &  & 0.053 & 1.176 & 0.054 & 1.179 & 0.041 & 1.196 & 0.041 & 1.196 \\ 
15.8 & 19 &  &  & 0.067 & 1.163 & 0.067 & 1.166 & 0.103 & 1.142 & 0.103 & 1.142 \\ 
23 & 32 & 0.087 & 1.114 &  &  &  &  &  &  &  &  \\
30 & 45 &  &  & 0.07 & 1.168 & 0.070 & 1.168 & 0.142 & 1.112 & 0.136 & 1.117 \\ 
52.8 & 91 &  &  & 0.080 & 1.159 & 0.080 & 1.159 & 0.170 & 1.090 & 0.170 & 1.090 \\ 
69 & 127.5 & 0.108 & 1.123 &  &  &  &  &  &  &  &  \\ 
85 & 164 &  &  & 0.094 & 1.146 & 0.094 & 1.145 & 0.187 & 1.083 & 0.187 & 1.083 \\ 
128.9 & 267 &  &  & 0.120 & 1.109 & 0.120 & 1.109 & 0.207 & 1.064 & 0.207 & 1.064 \\ 
157 & 341 & 0.114 & 1.118 &  &  &  &  &  & &  &  \\ 
186.4 & 415 &  &  & 0.118 & 1.119 & 0.118 & 1.119 & 0.208 & 1.071 & 0.208 & 1.071 \\
260.5 & 625 &  &  & 0.120 & 1.115 & 0.120 & 1.116 & 0.203 & 1.076 & 0.203 & 1.076 \\ 
308 & 773 & 0.119 & 1.115 &  &  &  &  &  &  &  &  \\ 
329.7 & 828 &  &  & 0.139 & 1.088 & 0.139 & 1.088 & 0.193 & 1.087 & 0.193 & 1.084 \\ 
382.8 & 1014 &  &  & 0.141 & 1.091 & 0.141 & 1.091 & 0.194 & 1.085 & 0.194 & 1.085 \\ 
\noalign{\smallskip}\hline\noalign{\smallskip}
\end{tabular}
\end{table*}

\clearpage

\newpage

\begin{figure*}[h]
\SetFigLayout{2}{2} \centering
\subfigure[]{\includegraphics[width=8cm]{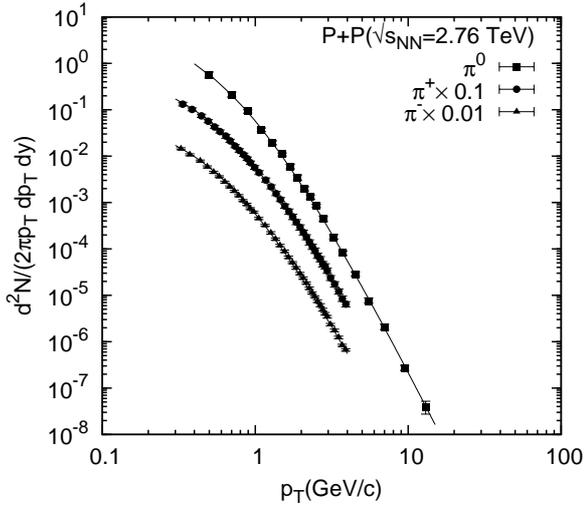}} \hfill
\subfigure[]{\includegraphics[width=8cm]{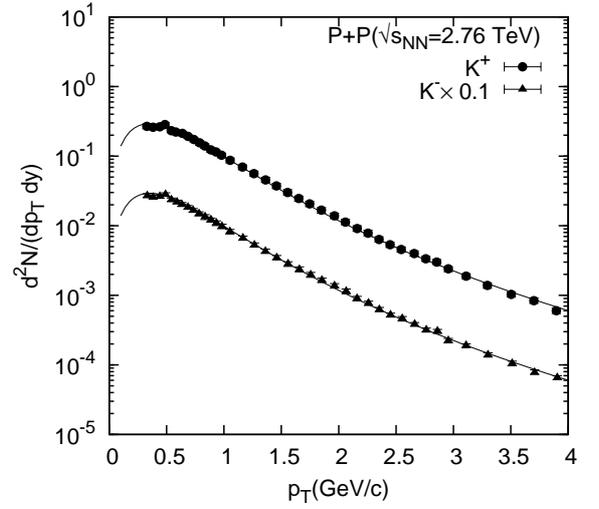}}
\caption{Plots of transverse momentum spectra of $\pi$- \& $K$-mesons
produced in $P+P$ collisions at $\sqrt{s_{NN}}=2.76$ TeV. The filled symbols represent the experimental data
points\cite{Peresunko1,Guerzoni1}. The solid curves provide the fits
on the basis of nonextensive approach(eqn.(10)).}
\end{figure*}

\begin{figure*}[h]
\SetFigLayout{2}{2} \centering
\subfigure[]{\includegraphics[width=8cm]{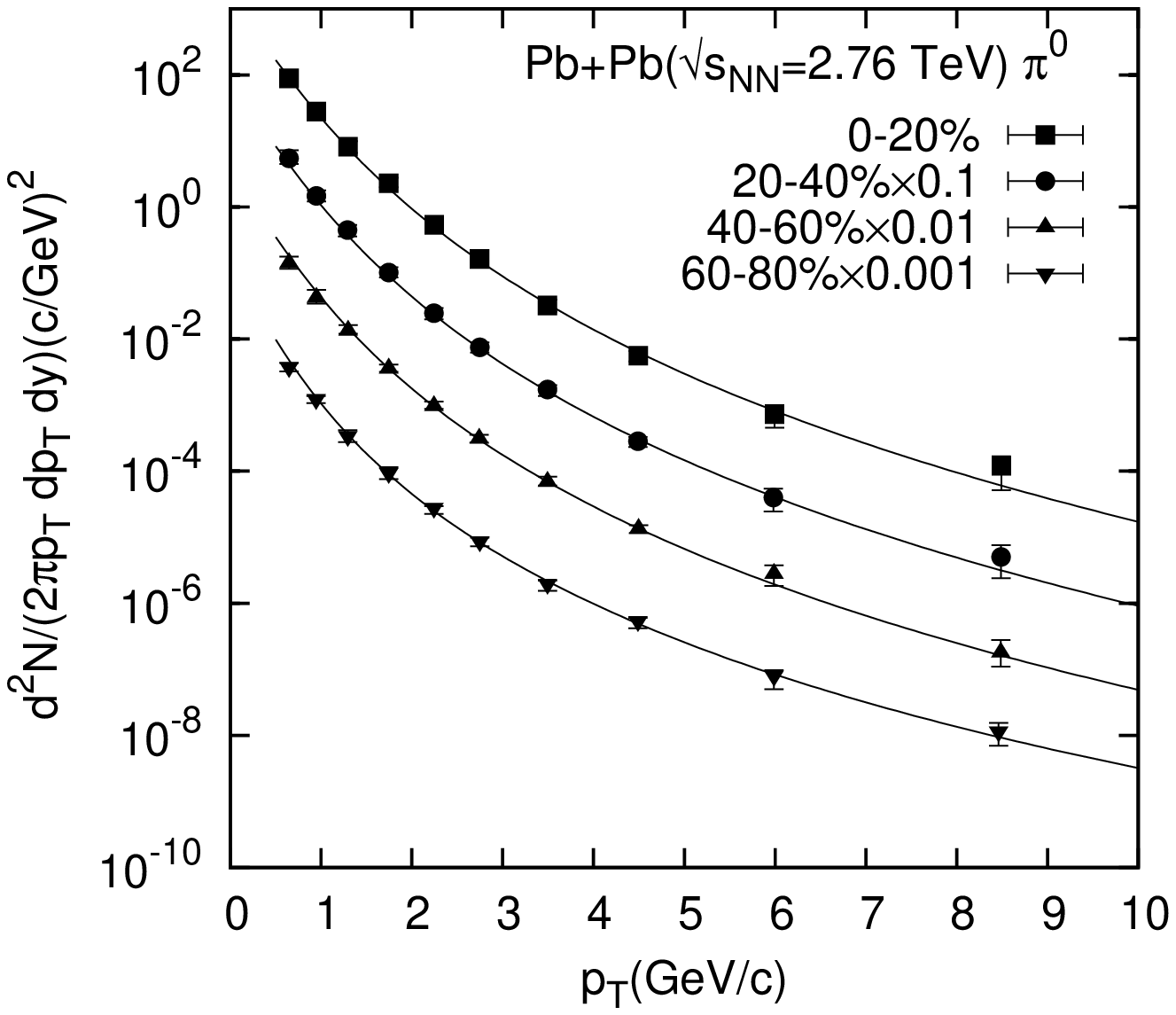}}\\
\subfigure[]{\includegraphics[width=8cm]{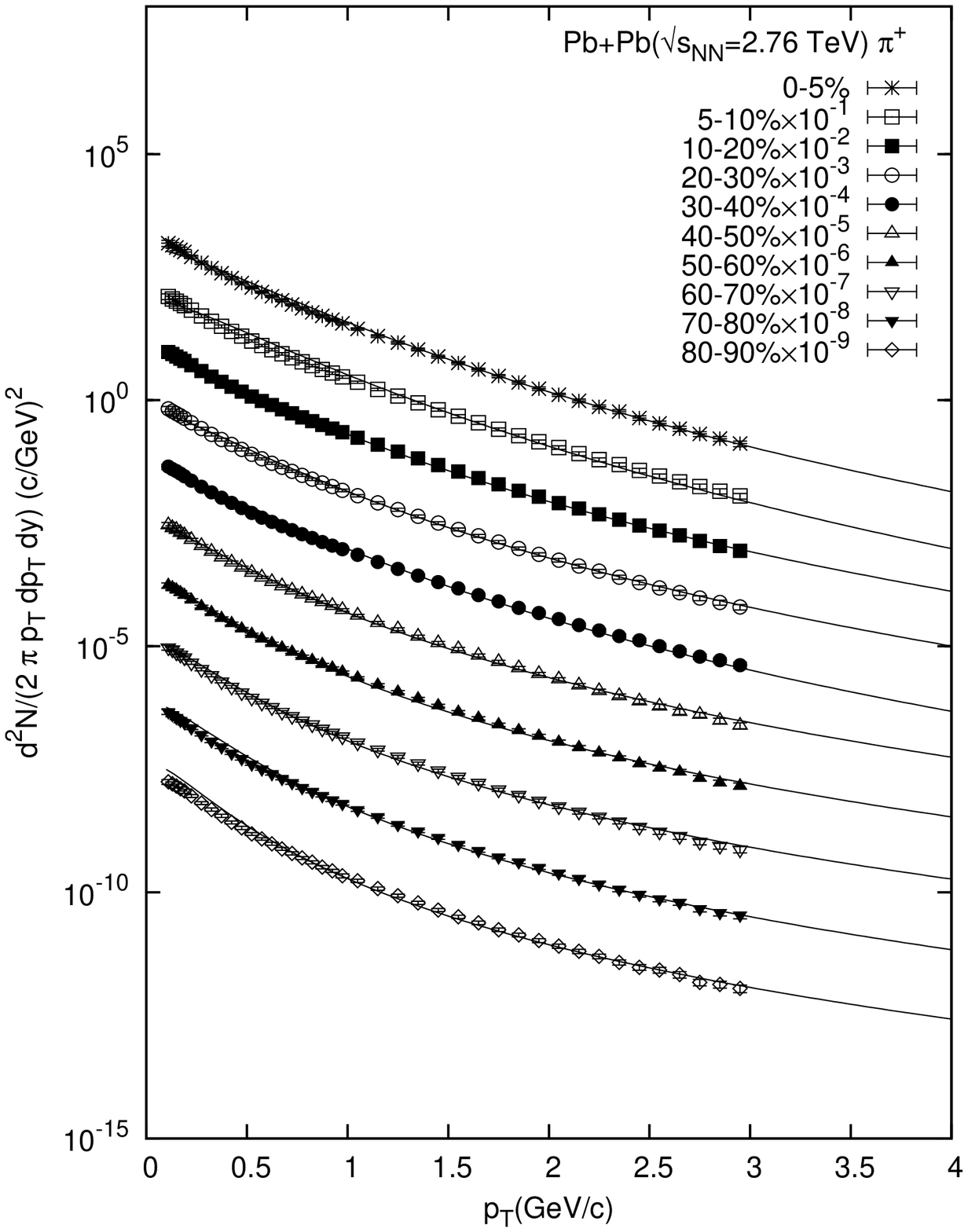}} 
\hfill
\subfigure[]{\includegraphics[width=8cm]{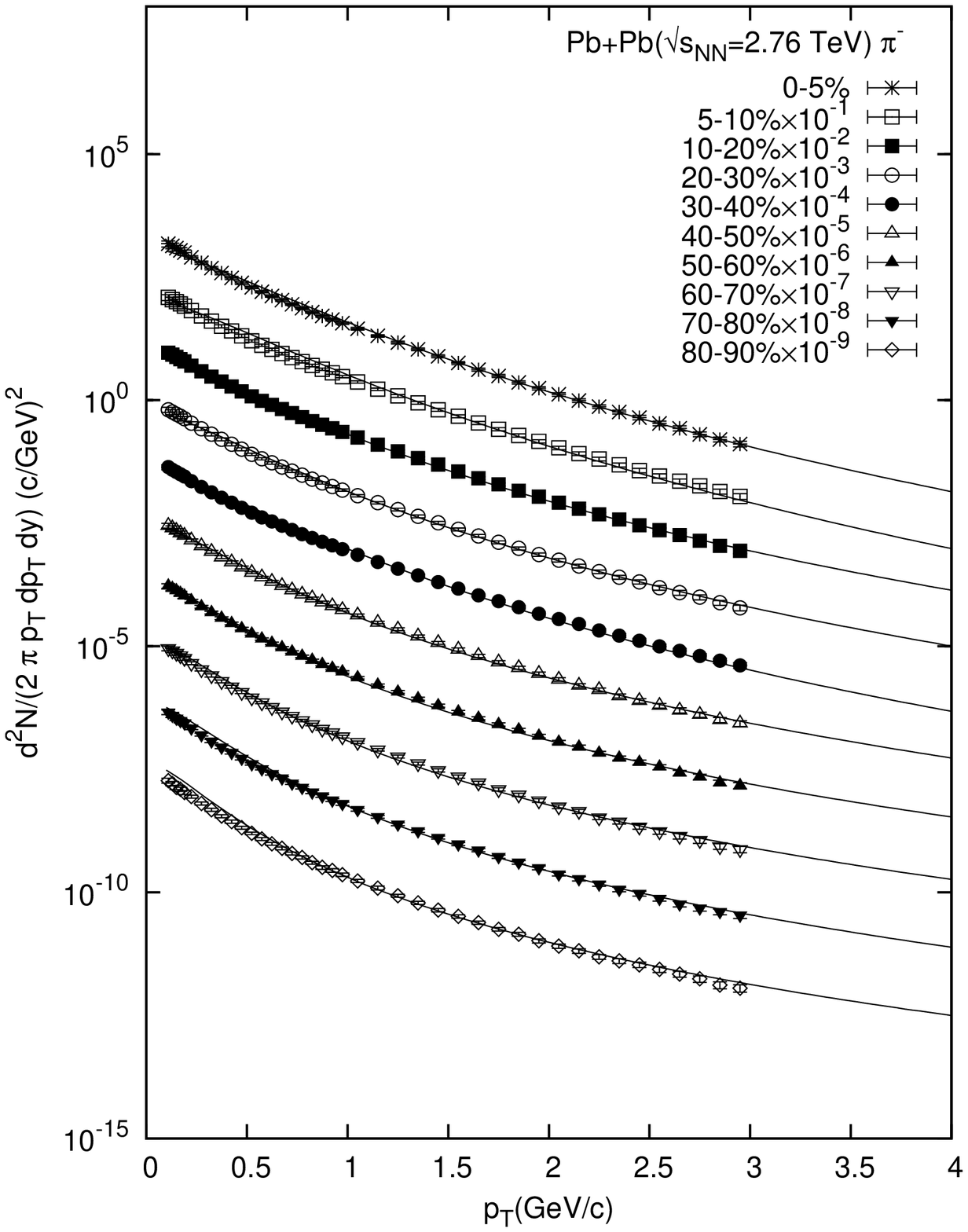}} 
\caption{Plots of Invariant yield of $\pi$-mesons
produced in different central $Pb+Pb$ collisions at $\sqrt{s_{NN}}=2.76$ TeV. The symbols represent the experimental data points\cite{Balbastre1,Abelev4} while the solid curves provide the fits
on the basis of nonextensive approach(eqn.(10-12)).}
\end{figure*}

\begin{figure*}[h]
\SetFigLayout{2}{2} \centering
\subfigure[]{\includegraphics[width=8cm]{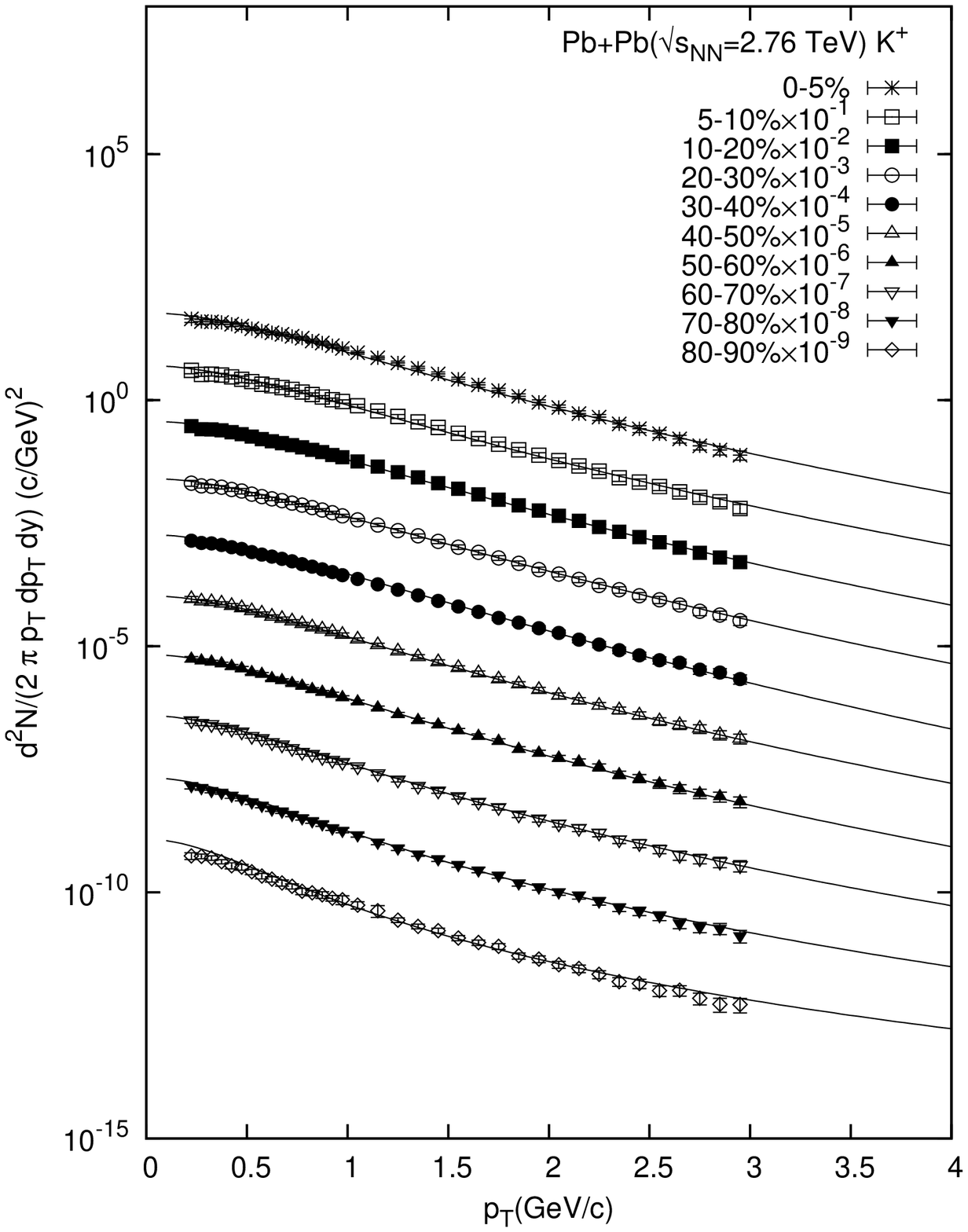}} 
\hfill
\subfigure[]{\includegraphics[width=8cm]{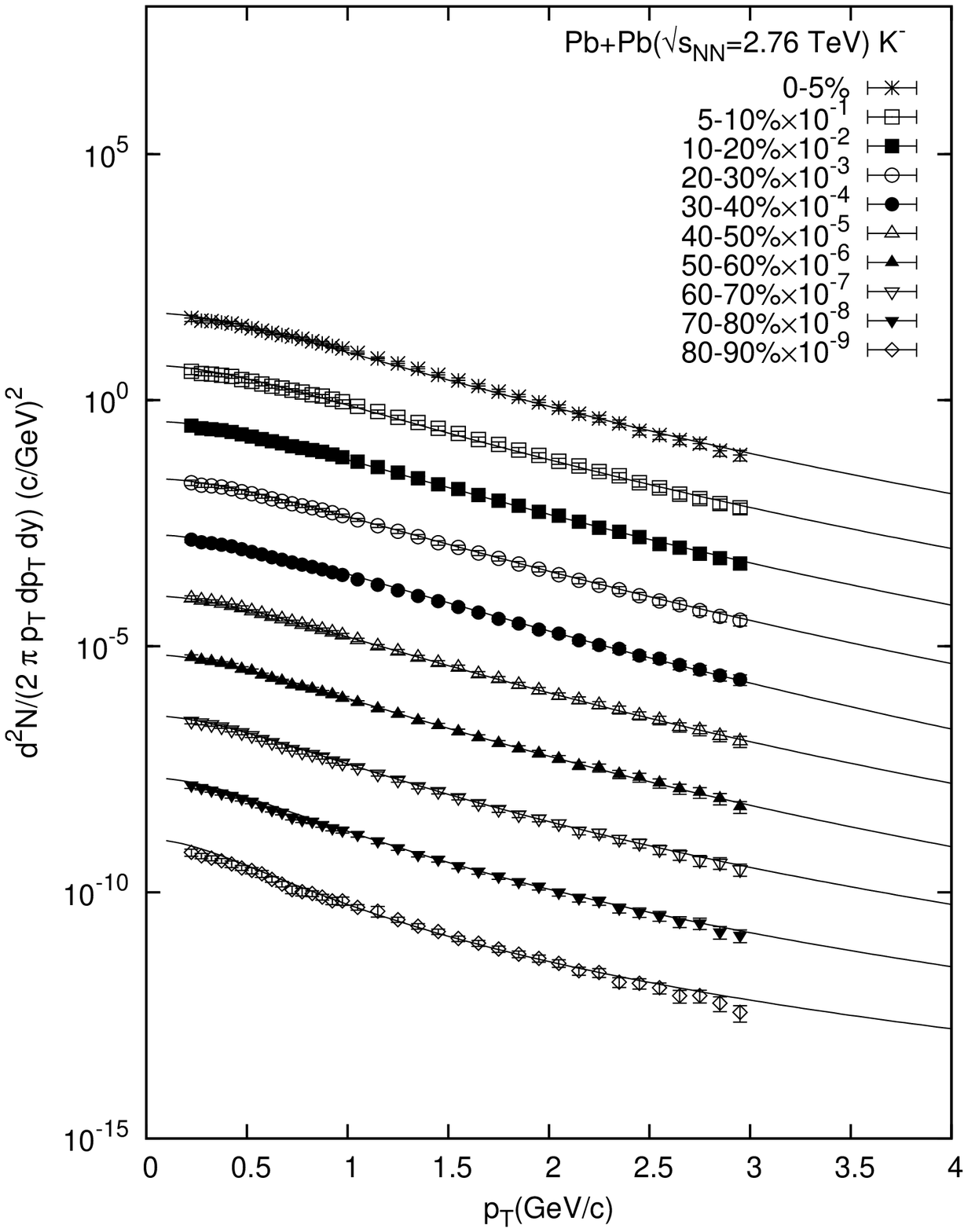}} 
\caption{Plots of Invariant yield of $K$-mesons
produced in different central $Pb+Pb$ collisions at $\sqrt{s_{NN}}=2.76$ TeV. The symbols represent the experimental data points\cite{Abelev4} and the solid curves provide the fits
on the basis of nonextensive approach(eqn.(10-12)).}
\end{figure*}
\begin{figure*}[h]
\centering
\includegraphics[width=8cm]{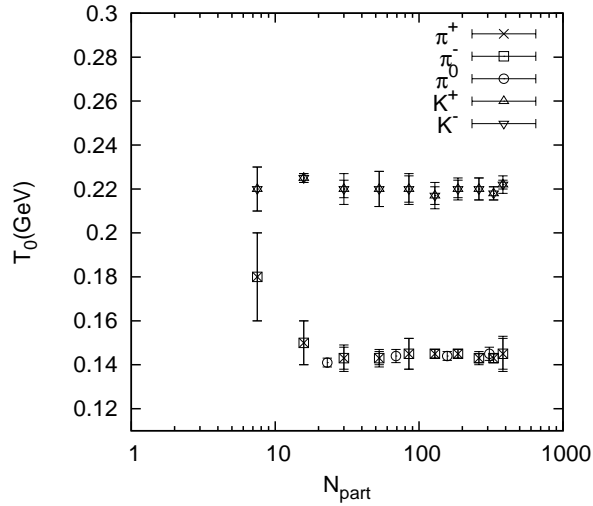}
\caption{Plot of the Hagedorn's Temperature $T_0$ obtained from the fits of different meson-spectra for different centralities of $Pb+Pb$ collisions at $\sqrt{s_{NN}}=2.76$ TeV. }
\end{figure*}
\begin{figure*}[h]
\SetFigLayout{1}{2} \centering
\subfigure[]{\includegraphics[width=8cm]{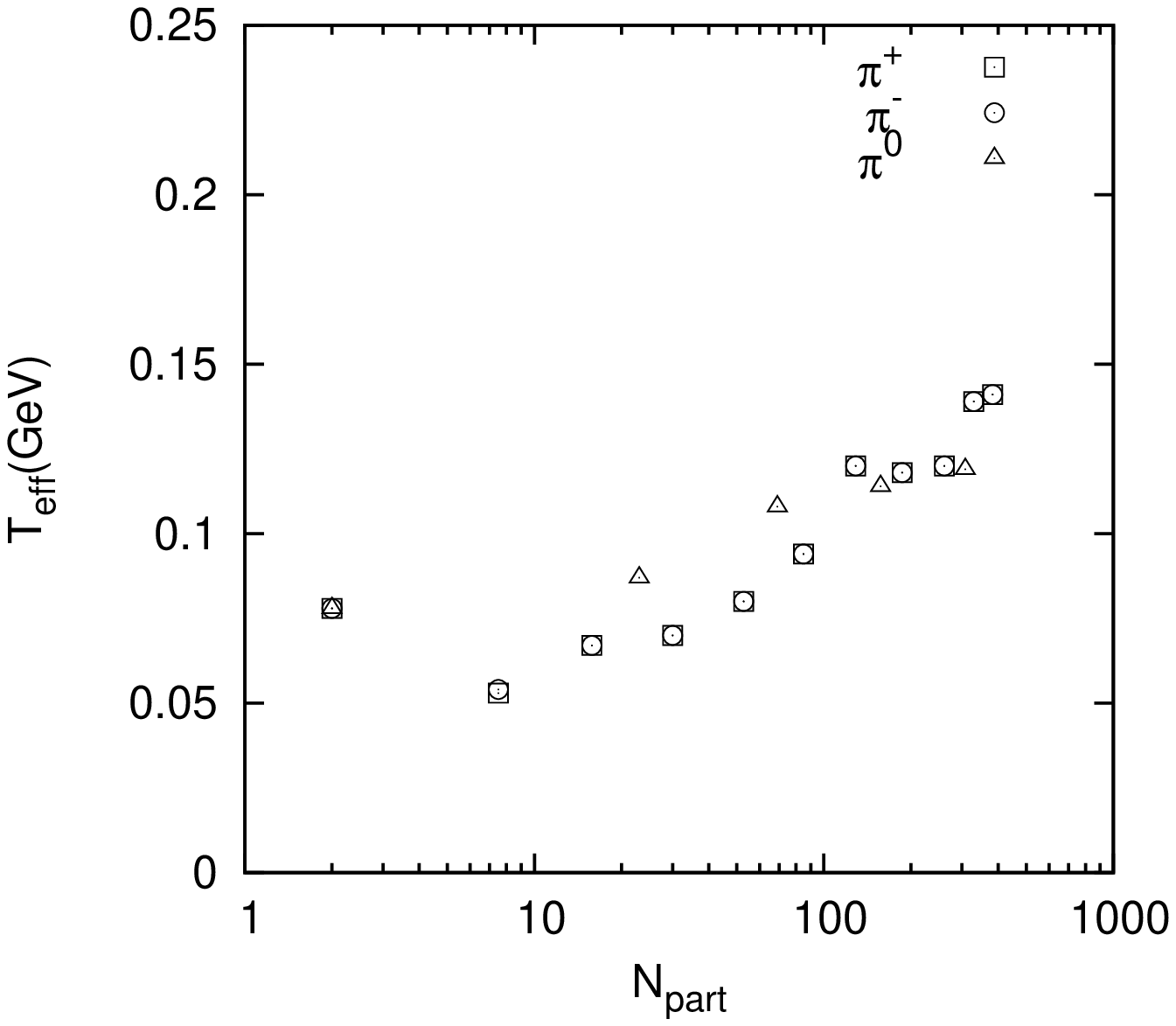}} \hfill
\subfigure[]{\includegraphics[width=8cm]{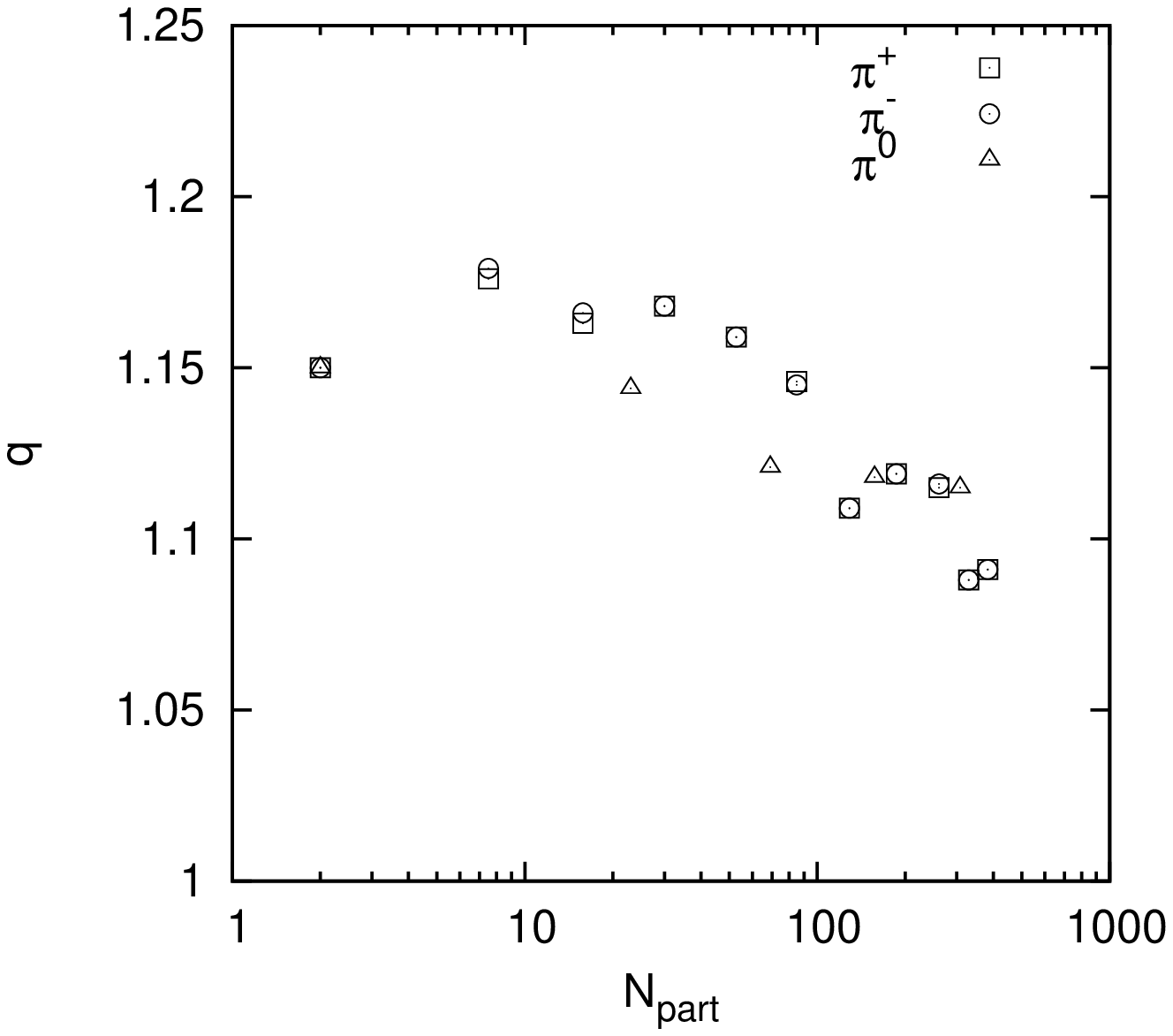}}\\
\subfigure[]{\includegraphics[width=8cm]{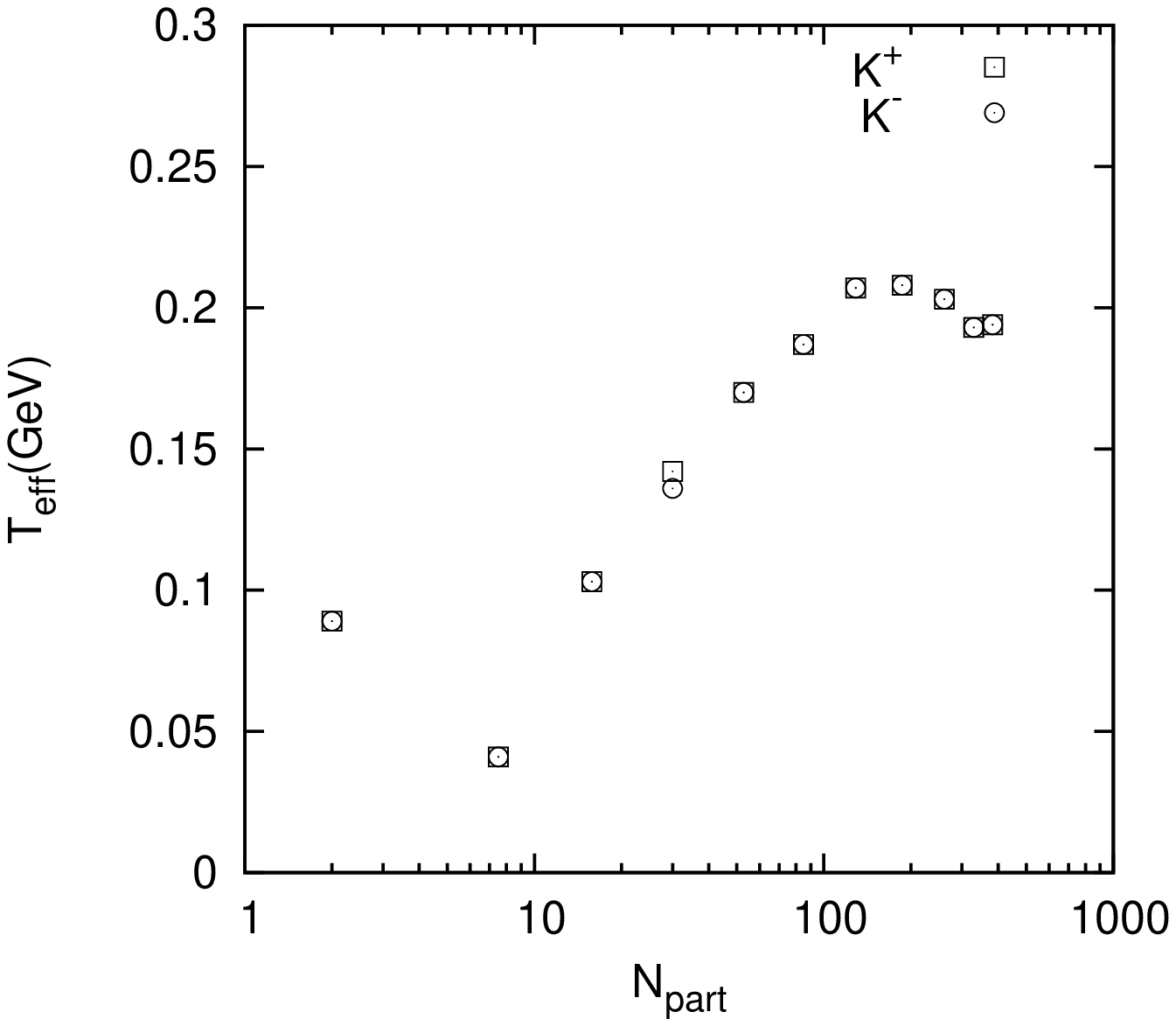}} \hfill
\subfigure[]{\includegraphics[width=8cm]{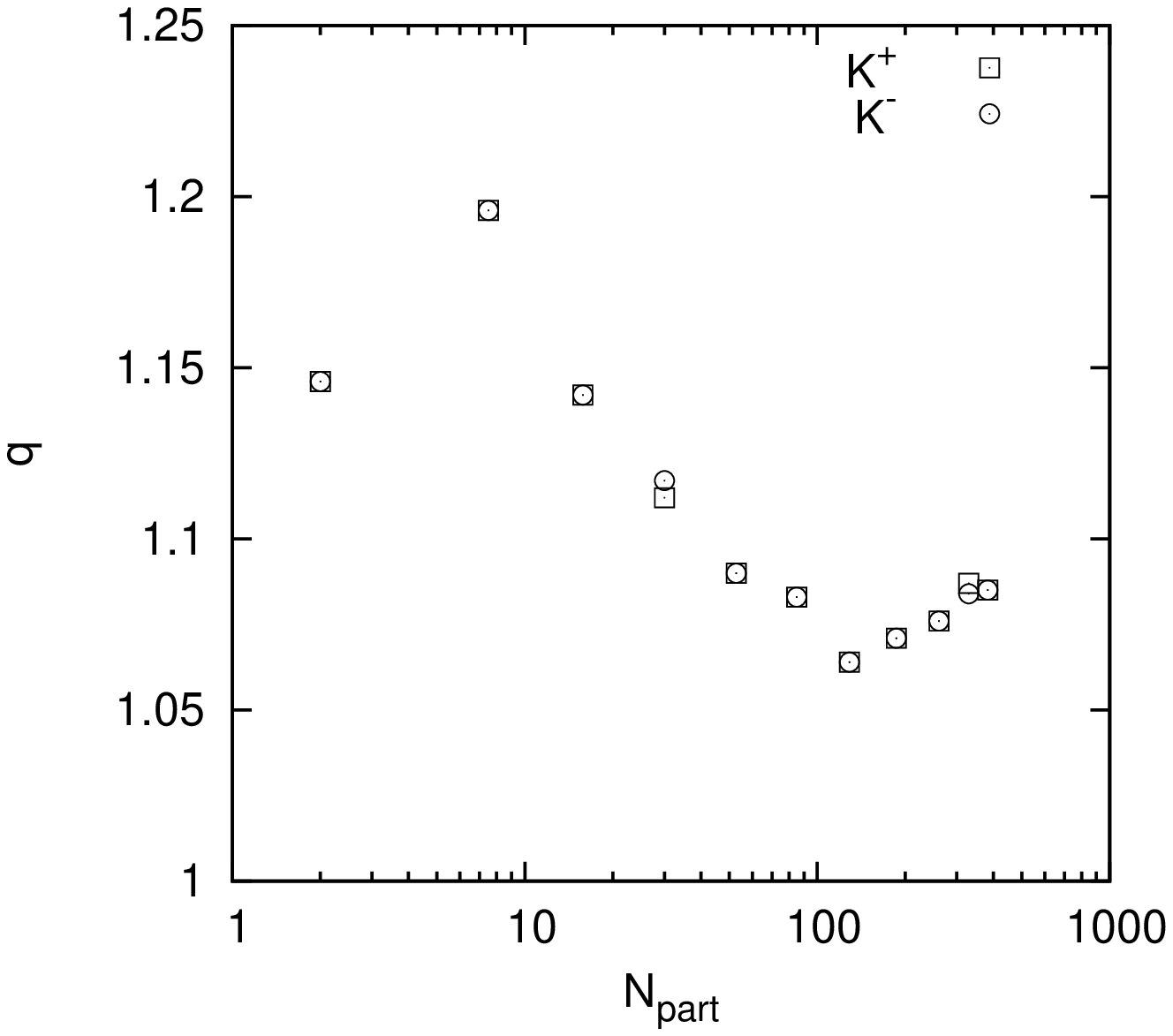}}
\caption{Plots of the effective temperature
$T_{eff}$  and the nonextensive parameter $q$  as a function of number of participant nucleons in $P+P$ and
$Pb+Pb$ collisions at $\sqrt{s_{NN}}=2.76$ TeV for production of secondary $\pi$ and K-mesons.}
\end{figure*}
\begin{figure*}[h]
\SetFigLayout{1}{2} \centering
\subfigure[]{\includegraphics[width=8cm]{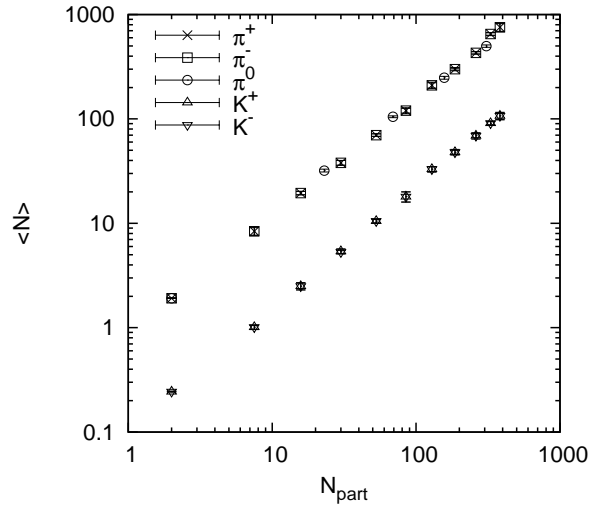}} \\
\subfigure[]{\includegraphics[width=8cm]{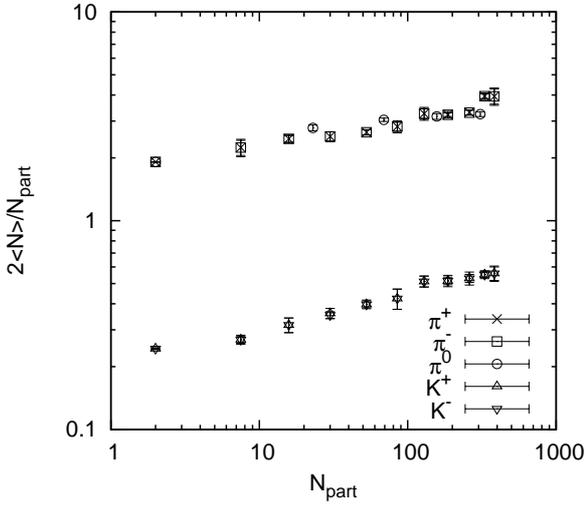}} \hfill
\subfigure[]{\includegraphics[width=8cm]{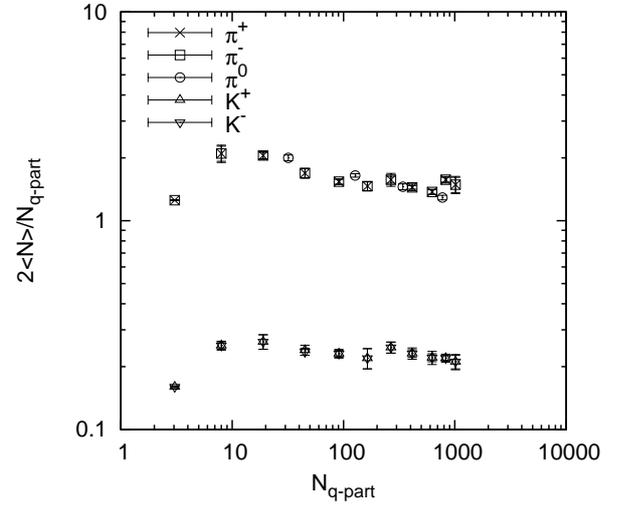}}
\caption{Plots of the average multiplicity for $\pi$ and K-mesons produced in $P+P$ and $Pb+Pb$ collisions at LHC energy $\sqrt{s_{NN}}=2.76$ TeV; and the same while normalized by pair of participant nucleons and pair of participant quarks respectively.}
\end{figure*}
\begin{figure*}[h]
\SetFigLayout{1}{2} \centering
\subfigure[]{\includegraphics[width=8cm]{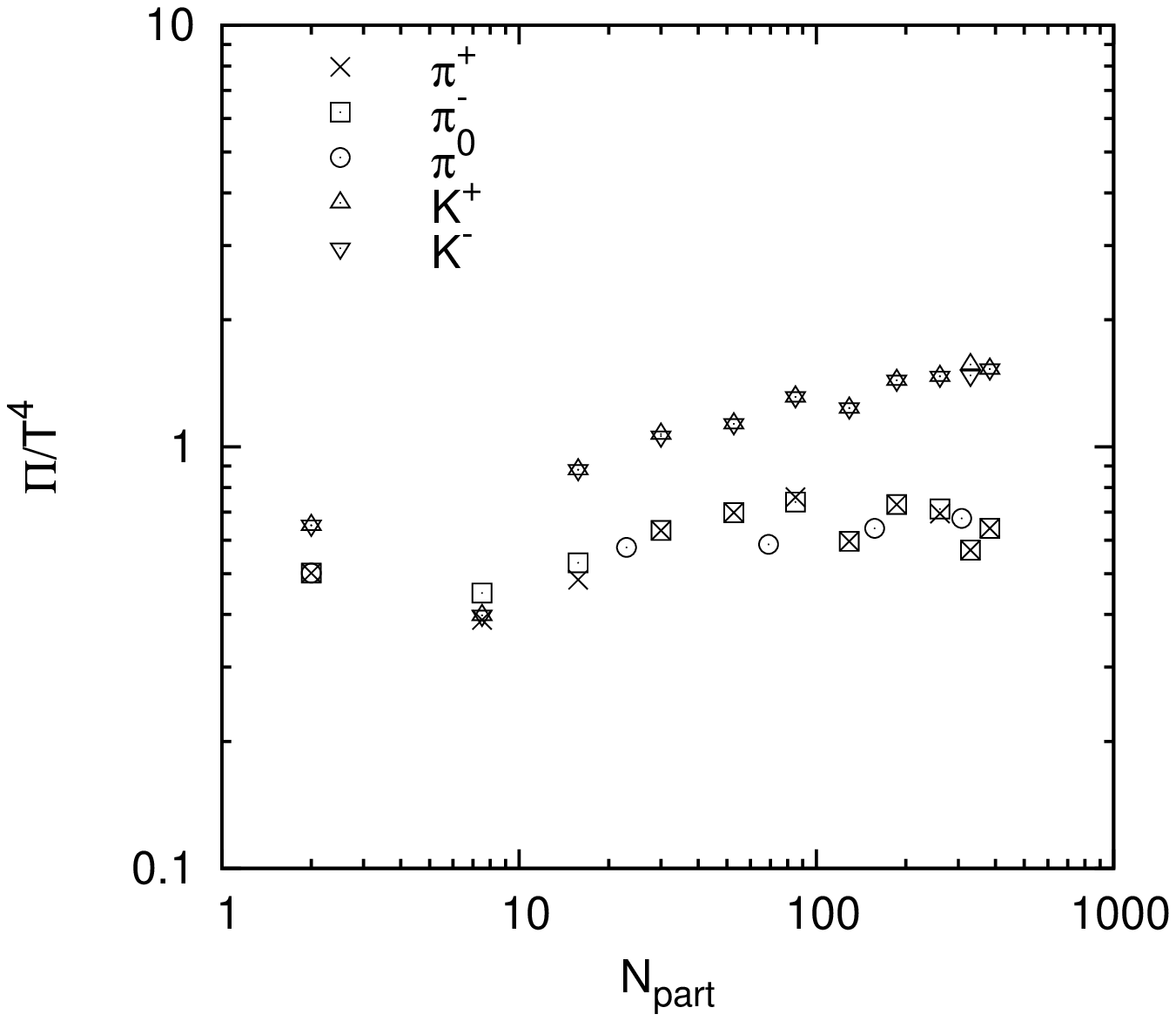}} \hfill
\subfigure[]{\includegraphics[width=8cm]{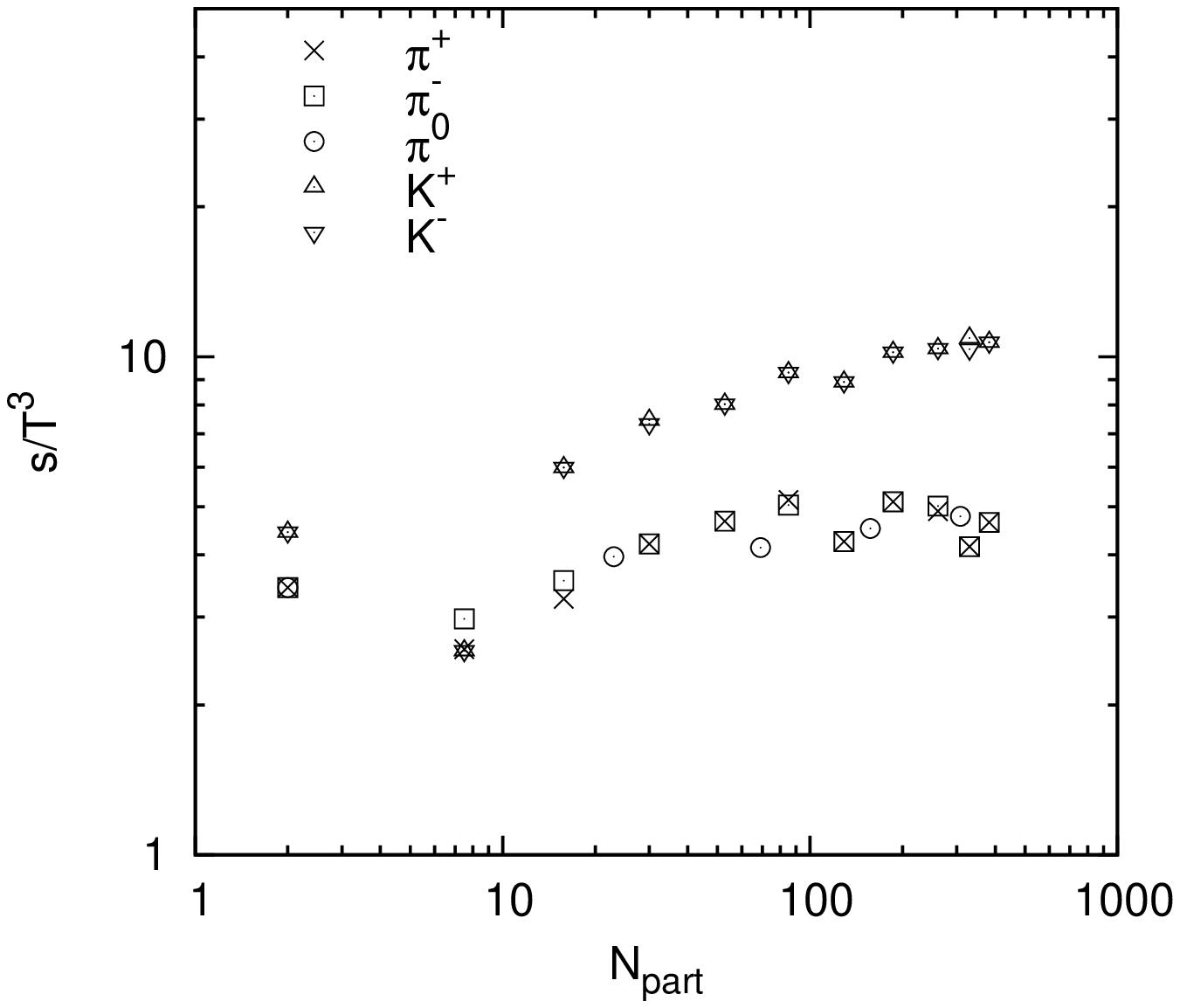}}\\
\subfigure[]{\includegraphics[width=8cm]{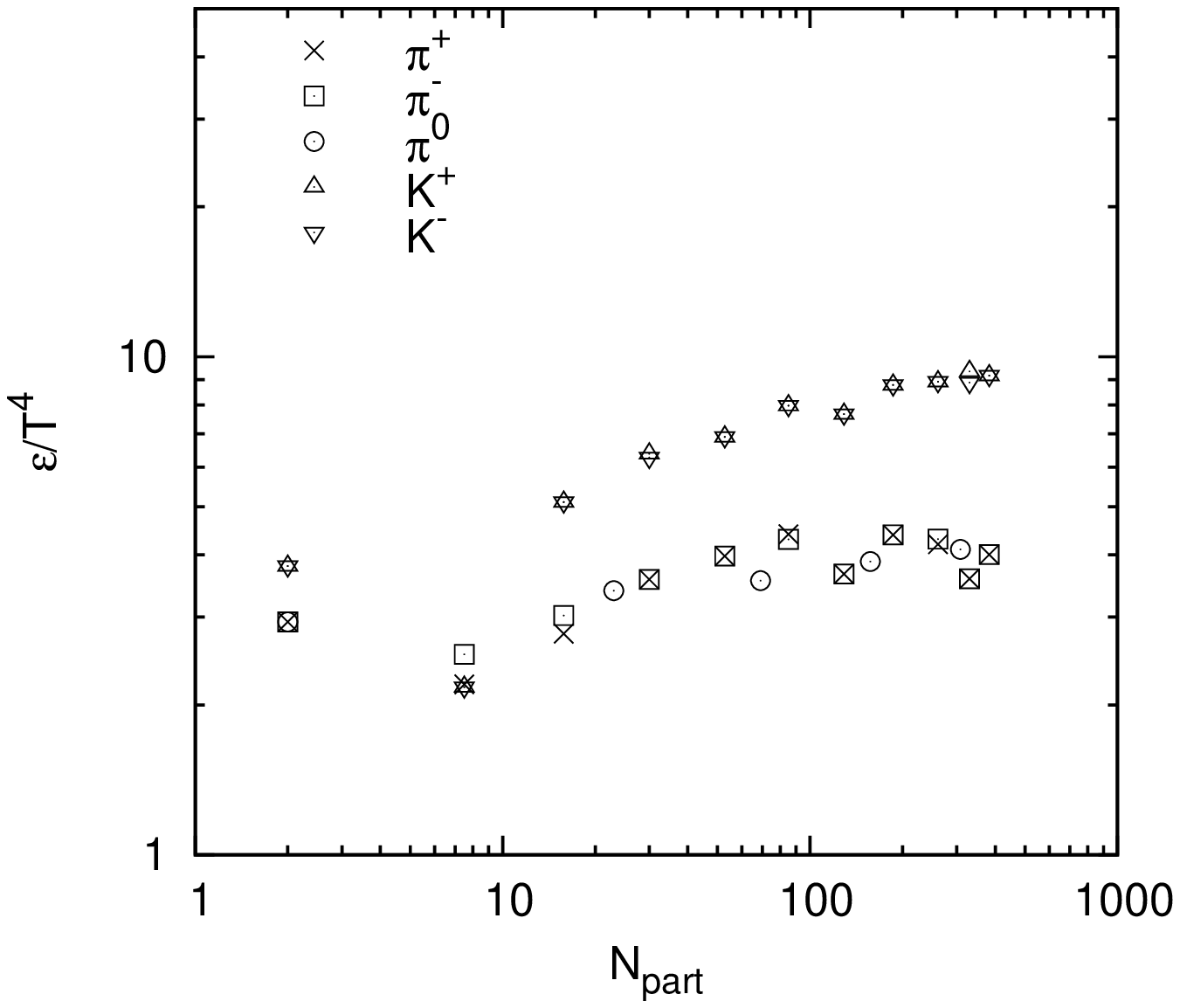}} \hfill
\subfigure[]{\includegraphics[width=8cm]{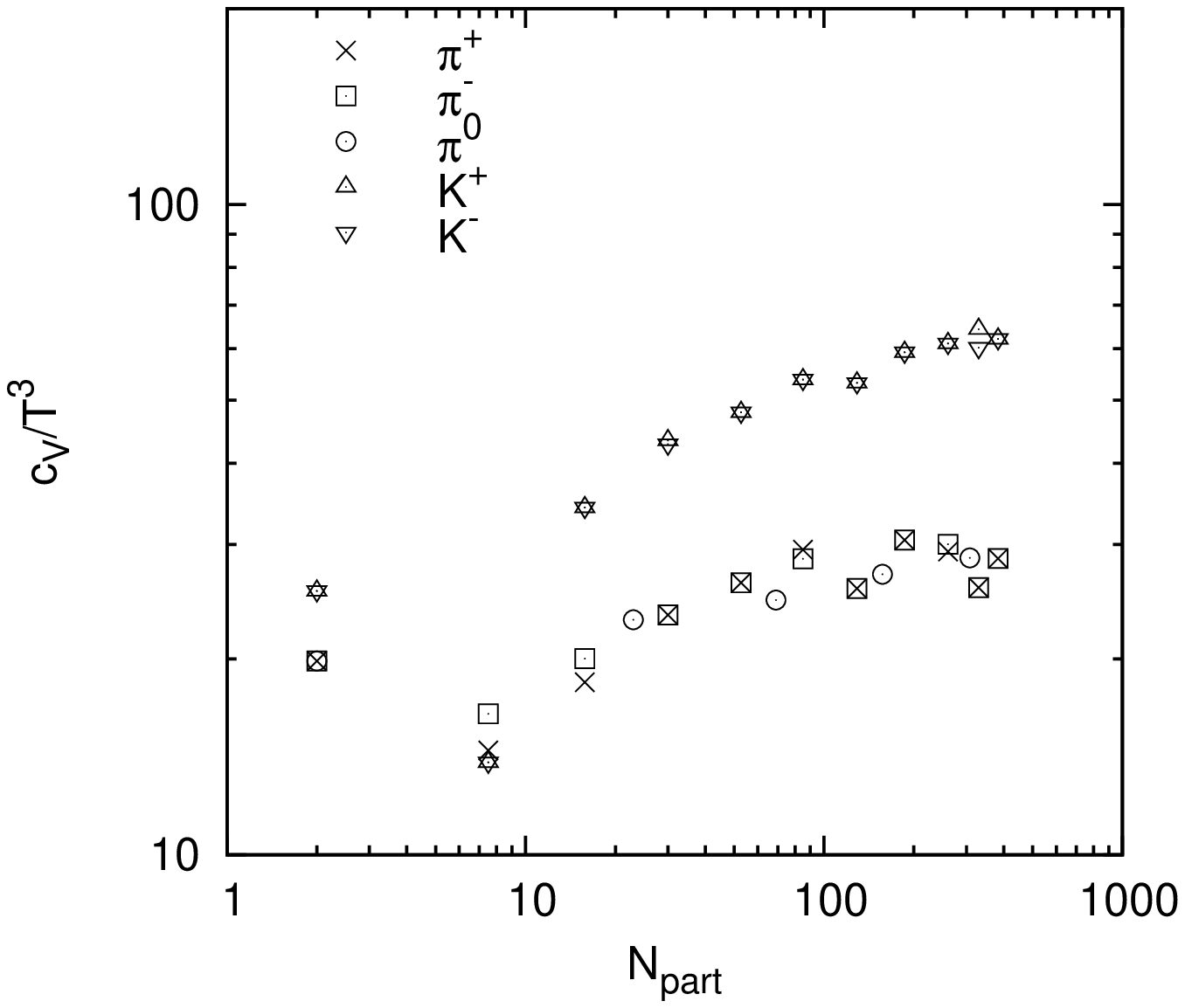}}
\caption{Plots of the Pressure($\Pi$, in units of $T^4$), entropy density($s$, in units of $T^3$), energy density($\epsilon$, in units of $T^4$) and specific heat density at constant volume($c_V$, in units of $T^3$) calculated for $P+P$ and different central $Pb+Pb$ collisions at LHC energy $\sqrt{s_{NN}}=2.76$ TeV.}
\end{figure*}
\begin{figure*}[h]
\SetFigLayout{1}{2} \centering
\subfigure[]{\includegraphics[width=8cm]{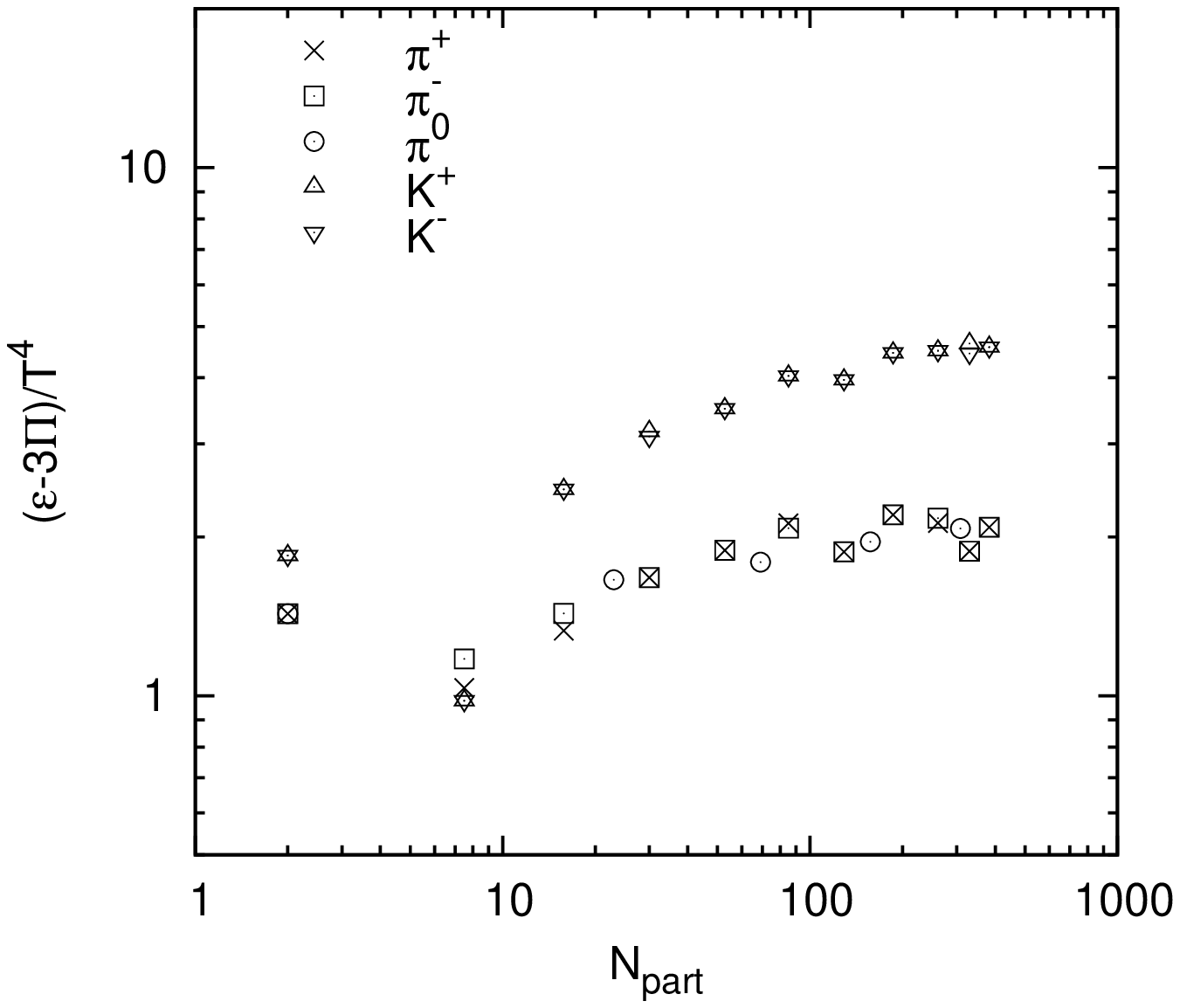}} \hfill
\subfigure[]{\includegraphics[width=8cm]{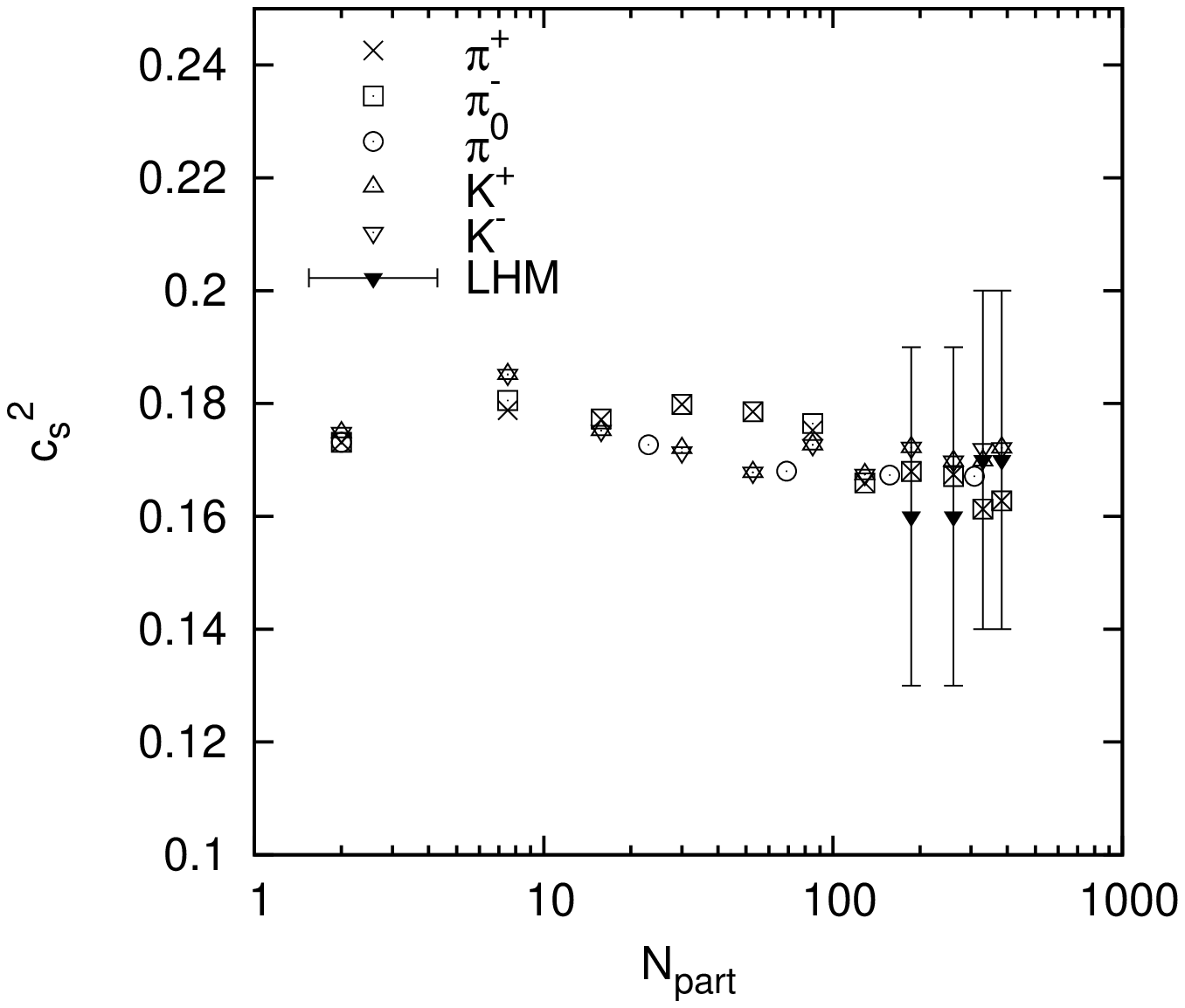}}
\caption{Plots of the trace anomaly($\frac{\epsilon-3\Pi}{T^2}$, in units of $T^2$) and velocity of sound squared($c_s^2$) calculated for $P+P$ and different central $Pb+Pb$ collisions at LHC energy $\sqrt{s_{NN}}=2.76$ TeV. The data points with error-bars in (b) are the results on the basis of a Landau Hydrodynamic model\cite{Gao1}.}
\end{figure*}
\begin{figure*}[h]
\SetFigLayout{1}{2} \centering
\subfigure[]{\includegraphics[width=8cm]{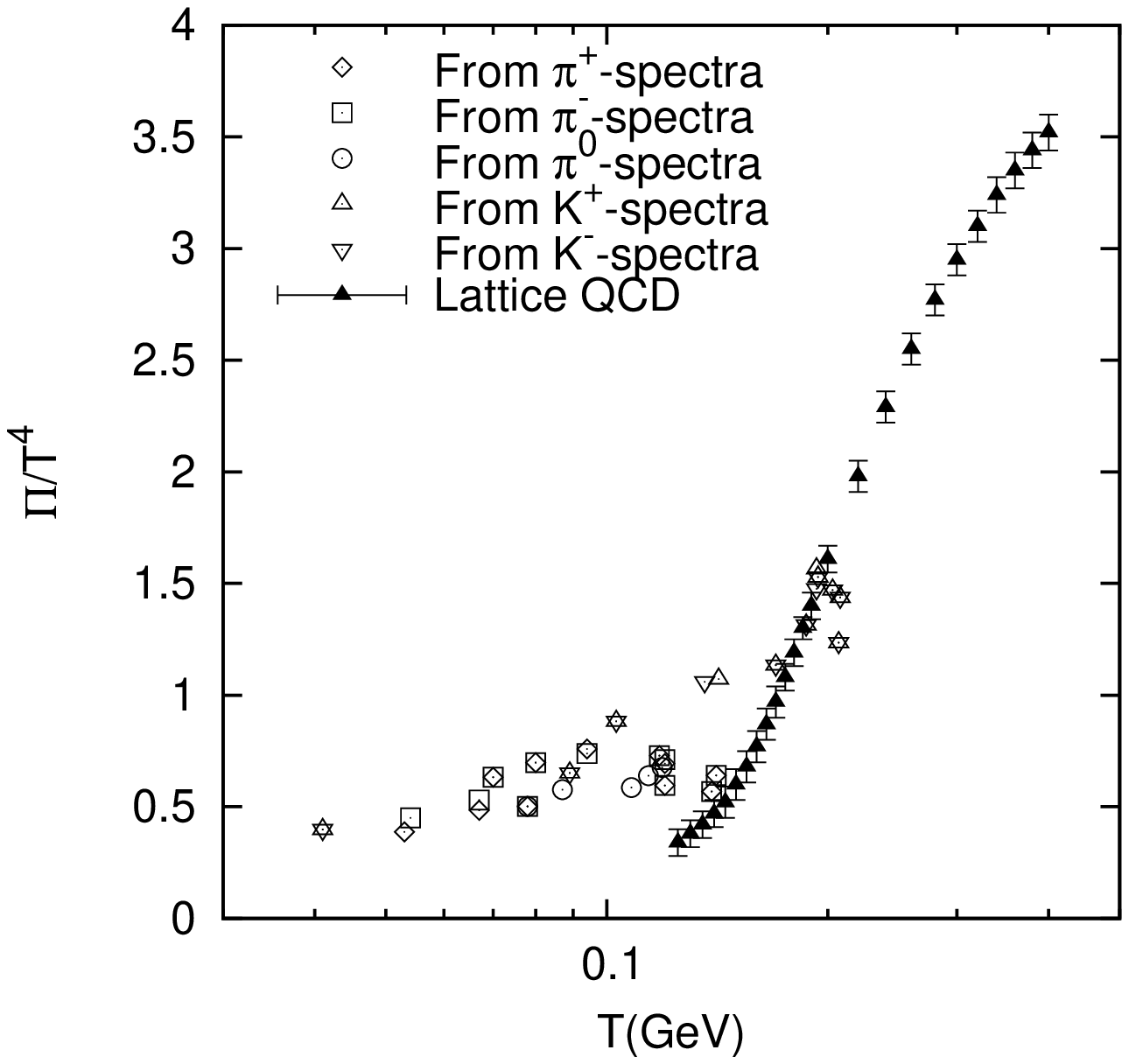}} \hfill
\subfigure[]{\includegraphics[width=8cm]{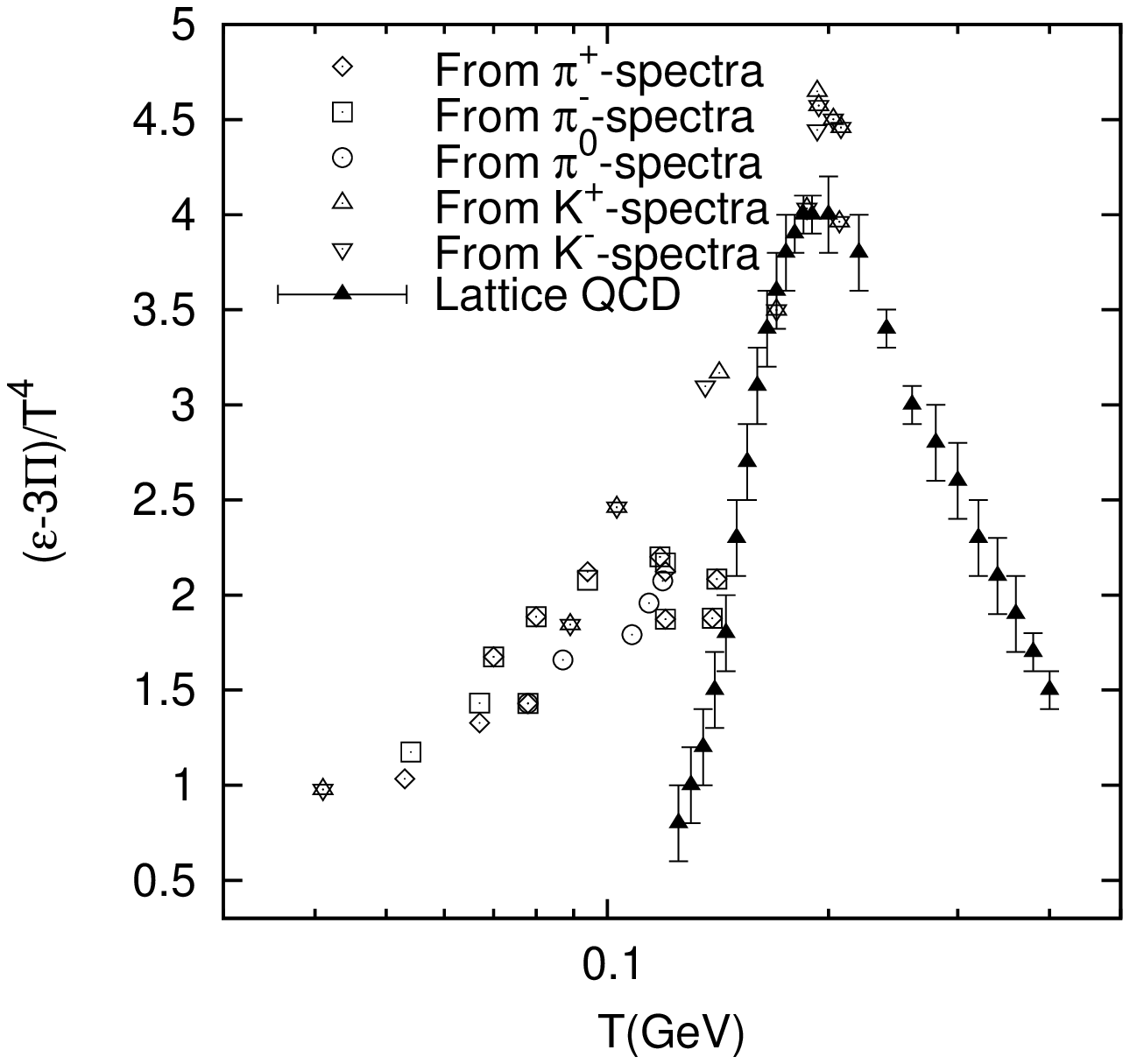}}\\
\subfigure[]{\includegraphics[width=8cm]{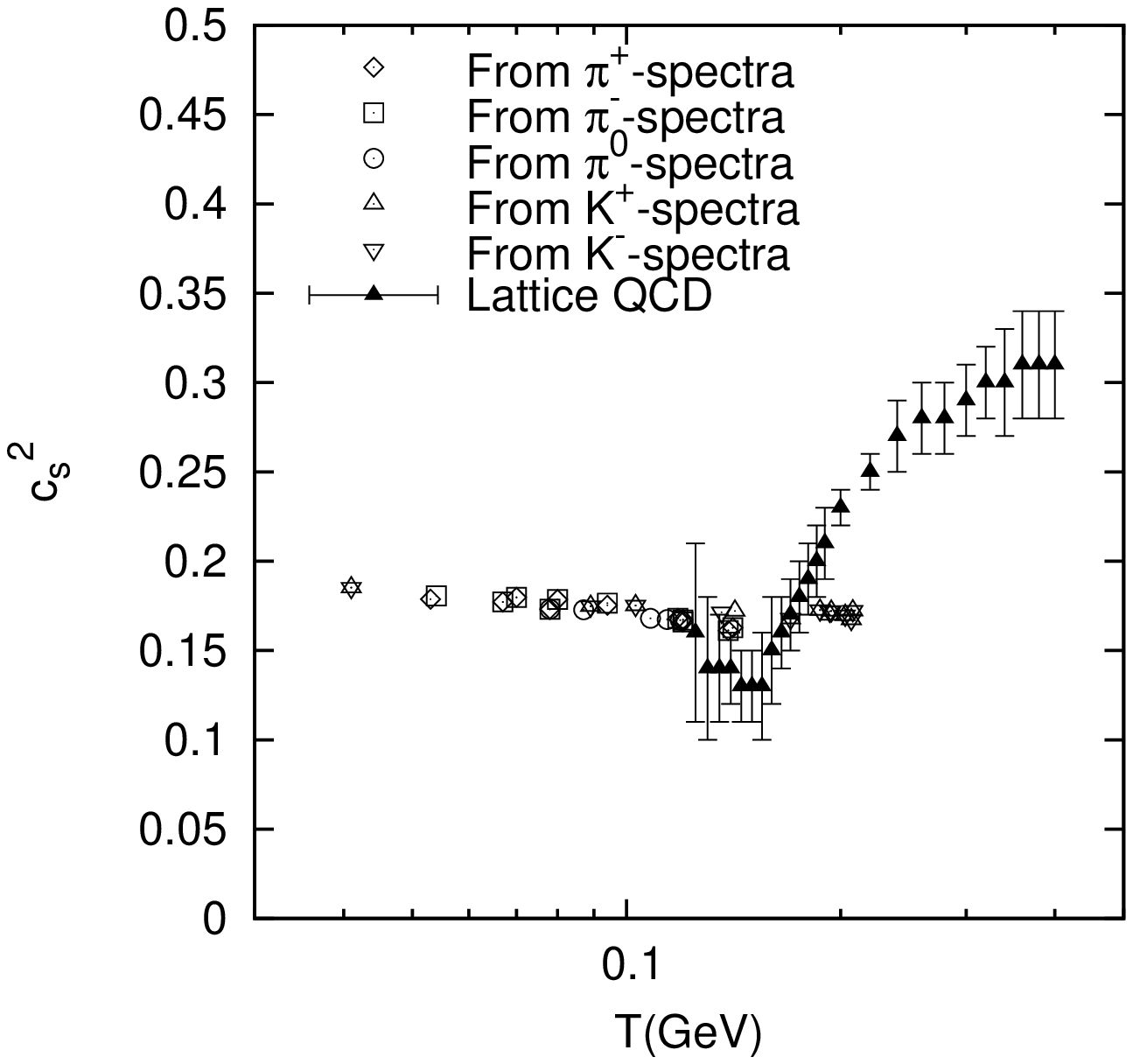}}
\caption{Comparison of some thermal parameters obtained in the present work with those from a Lattice QCD calculation\cite{Borsanyi1}.}
\end{figure*}

\end{document}